\documentclass[preprint]{aastex}
\newcommand\pcc{\;{\rm cm}^{-3}}
\newcommand\psc{\;{\rm cm}^{-2}}
\newcommand\Msun{\; {\rm M}_{\odot}}
\newcommand\kms{\; {\rm km}\;{\rm s}^{-1}}
\newcommand\ergs{\; {\rm erg}\;{\rm s}^{-1}}

\newcommand\cm{\;{\rm cm}}
\newcommand\yr{\; {\rm yr}}

\newcommand\pc{\;{\rm pc}}
\newcommand\kpc{\;{\rm kpc}}
\newcommand\sfrunit{\Msun \kpc^{-2} \yr^{-1}}
\newcommand\Punit{\pcc\,{\rm K}}
\newcommand\Surf{\Msun\;{\rm pc^{-2}}}
\newcommand\rhounit{\Msun\;{\rm pc^{-3}}}
\newcommand\Kel{\;{\rm K}}

\newcommand\simgt{\lower.5ex\hbox{$\; \buildrel > \over \sim \;$}}
\newcommand\simlt{\lower.5ex\hbox{$\; \buildrel < \over \sim \;$}}

\newcommand\rbrackets[1]{\left({#1}\right)}
\newcommand\sbrackets[1]{\left[{#1}\right]}

\newcommand\Pth{P_{\rm th}}
\newcommand\Ptwo{P_{\rm two}}

\newcommand\Pturb{P_{\rm turb}}
\newcommand\Pdriv{P_{\rm driv}}

\newcommand\Pmax{P_{\rm max}}
\newcommand\Pmin{P_{\rm min}}

\newcommand\kbol{k_{\rm B}}
\newcommand\rhosd{\rho_{\rm sd}}
\newcommand\tauint{\tau_{\rm int}}
\newcommand\Nsim{N_{\rm H}}
\newcommand\Nsyn{N_{\rm H,obs}}
\newcommand\Nthin{N_{\rm H,thin}}
\newcommand\Nth{N_{\rm H,lim}}

\newcommand\Tspeak{T_{s, {\rm peak}}}
\newcommand\taupeak{\tau_{\rm peak}}
\newcommand\vfwhm{\Delta v_{\rm FWHM}}

\newcommand\Tsim{T_{s, {\rm avg}}}
\newcommand\Tsyn{T_{s, {\rm obs}}}
\newcommand\Tsavg{T_{s, {\rm avg}}(\vch)}
\newcommand\Tkavg{T_{k, {\rm avg}}(\vch)}
\newcommand\Tsobs{T_{s, {\rm obs}}(\vch)}
\newcommand\fcsyn{f_{c, {\rm obs}}}
\newcommand\fcsim{f_{c}}
\newcommand\vch{v_{\rm ch}}
\newcommand\Nwnm{N_{\rm WNM}}

\newcommand\smax{s_{\rm max}}
\newcommand\torb{t_{\rm orb}}
\newcommand\tosc{t_{\rm osc}}
\newcommand\SigSFR{\Sigma_{\rm SFR}}

\shorttitle{Synthetic HI Observations}
\shortauthors{Kim, Ostriker, \& Kim}

\begin{document}

\title{Three Dimensional Hydrodynamic Simulations of Multiphase Galactic Disks
with Star Formation Feedback: II. Synthetic \ion{H}{1} 21~cm Line Observations}

\author{Chang-Goo Kim\altaffilmark{1}, 
Eve C.\ Ostriker\altaffilmark{1}, 
and Woong-Tae Kim\altaffilmark{2}}

\affil{$^1$Department of Astrophysical Sciences, Princeton University,
Princeton, NJ, 08544, USA} 
\email{cgkim@astro.princeton.edu, eco@astro.princeton.edu}
\affil{$^2$Center for the Exploration of the Origin of the Universe (CEOU),
Astronomy Program, Department of Physics \& Astronomy, Seoul National
University, Seoul 151-742, Republic of Korea}
\email{wkim@astro.snu.ac.kr}

\begin{abstract}
We use three-dimensional numerical hydrodynamic simulations of the turbulent,
multiphase atomic interstellar medium (ISM) to construct and analyze synthetic
\ion{H}{1} 21~cm emission and absorption lines.  Our analysis provides detailed
tests of 21~cm observables as physical diagnostics of the atomic ISM.  In
particular, we construct (1) the ``observed'' spin temperature, $\Tsobs\equiv
T_B(\vch)/[1-e^{-\tau(\vch)}]$, and its optical-depth weighted mean $\Tsyn$;
(2) the absorption-corrected ``observed'' column density, $\Nsyn \propto \int
d\vch  T_B(\vch) \tau(\vch)/[1-e^{-\tau(\vch)}]$; and (3) the ``observed''
fraction of cold neutral medium (CNM), $\fcsyn\equiv T_c/\Tsyn$ for $T_c$ the
CNM temperature; we compare each observed parameter with true values obtained
from line-of-sight (LOS) averages in the simulation.  Within individual
velocity channels, $\Tsobs$ is within a factor 1.5 of the true value up to
$\tau(\vch)\sim 10$.  As a consequence, $\Nsyn$ and $\Tsyn$ are respectively
within 5\% and 12\% of the true values for 90\% and 99\% of LOSs.  The
optically thin approximation significantly underestimates $\Nsim$ for $\tau>1$.
Provided that $T_c$ is constrained, an accurate observational estimate of the
CNM mass fraction can be obtained down to 20\%. 
 We show that $\Tsyn$ cannot be
used to distinguish the relative proportions of warm and thermally-unstable
atomic gas, although the presence of thermally-unstable gas can be discerned
from 21~cm lines with $200\Kel\simlt\Tsobs\simlt1000\Kel$.
Our mock observations successfully reproduce and explain the observed
distribution of the brightness temperature, optical depth, and spin temperature
in \citet{2013MNRAS.436.2352R}.  The threshold column density for CNM seen in
observations is also reproduced by our mock observations.  We explain this
observed threshold behavior in terms of vertical equilibrium in the local Milky
Way's ISM disk.
\end{abstract}


\section{INTRODUCTION}\label{sec:intro}

The \ion{H}{1} 21~cm line is a powerful tool for studying the atomic
interstellar medium (ISM).  The first detections of emission and absorption at
21~cm date back to the 1950s (\citealt{1951Natur.168..356E,1951Natur.168..357M}
for emission, and \citealt{1955ApJ...122..361H} for absorption). The \ion{H}{1}
21~cm line has been observed extensively since then, and proved extremely
valuable in revealing many properties of the Milky Way Galaxy, including the
vertical and radial distribution and warp of the atomic disk, the galactic
rotation curve and dark matter distribution, and spiral structure.  In addition
to large-scale properties of the Milky Way, \ion{H}{1} 21~cm observations
provide a wealth of knowledge regarding the detailed physical state of the
interstellar medium (see reviews including
\citealt{1976ARA&A..14..275B,1990ARA&A..28..215D,2009ARA&A..47...27K} and
references therein). Recently, the combined Leiden-Argentine-Bonn survey (LAB
survey, \citealt{2005A&A...440..775K}) produced a high-sensitivity 21~cm
emission map over the entire sky with 36 arcmin resolution by merging the
Leiden-Dwingeloo survey \citep{1997agnh.book.....H} with the Instituto
Argentino de Radioastronom\'ia survey \citep{2005A&A...440..767B}. This
high-sensitivity, single-dish survey with stray radiation correction enables
highly detailed investigation of \ion{H}{1} in the Milky Way
\citep{2006Sci...312.1773L,2006ApJ...643..881L,2007A&A...469..511K,2008A&A...487..951K}.

Information about the thermodynamic state of hydrogen can be obtained from
emission/absorption line pairs, which yield the excitation temperature (a.k.a.
spin temperature) of the 21~cm line.  Since the available radio continuum
background sources are generally weak at 21~cm, the line profile toward a
continuum source gives a mixture of emission and absorption by \ion{H}{1}. In
order to separate emission and absorption, \ion{H}{1} emission contributions
must be estimated from off-source measurements with sufficient spatial
resolution.  Using the largest single dish telescope, the Arecibo telescope
(beamwidth of 3.2 arcmin), \citet{2003ApJS..145..329H} have investigated the
emission/absorption line pairs toward continuum sources at high and
intermediate latitudes, but the resolution from a single dish is not sufficient
for accurate interpolated emission at latitudes below $10^\circ$.
Interferometric surveys including the Canadian Galactic Plane Survey (CGPS,
\citealt{2003AJ....125.3145T}), the Southern Galactic Plane Survey (SGPS,
\citealt{2005ApJS..158..178M}), and the VLA Galactic Plane Survey (VGPS,
\citealt{2006AJ....132.1158S}) achieve angular resolution about $\sim$1 arcmin.
In order to overcome inherent low sensitivity of interferometric observations
to extended structures (zero-spacing), these new surveys apply short-spacing
corrections from single dish observations, allowing accurate measurement of the
expected emission and hence absorption spectra \citep{2006AJ....132.1158S}. The
resulting emission/absorption line pairs in these Galactic plane surveys reveal
a pervasive multiphase structure in the atomic ISM out to a galactocentric
radius of $25\kpc$ \citep{2009ApJ...693.1250D}.

Using emission/absorption line pairs, the total atomic column density and the
harmonic mean spin temperature on a given line-of-sight (LOS) can be derived
with a simple radiative transfer calculation \citep[][see Section~\ref{sec:syn}
below]{1978ppim.book.....S,2011piim.book.....D}. In classical models of the
neutral ISM, two distinct phases are expected: the cold neutral medium (CNM;
$T\sim 100\Kel$) and the warm neutral medium (WNM; $T\sim 8000\Kel$), in
pressure equilibrium with each other
\citep[e.g.,][]{1969ApJ...155L.149F,1995ApJ...443..152W,2003ApJ...587..278W}.
Since the WNM is optically thin at 21~cm due to its low density and high
temperature, a simple single temperature approximation for radiative transfer
would be satisfied if there is no overlap between CNM clouds within narrow
velocity channels.  However, recent dynamical models that simulate the atomic
ISM
\citep[e.g.,][]{2000ApJ...540..271V,2005A&A...433....1A,2005ApJ...626..864M,2007A&A...465..431H,2007ApJ...663..183P,2008ApJ...681.1148K,2011A&A...526A..14S,2013arXiv1301.3446S},
along with detailed observations
\citep[e.g.,][]{2003ApJ...586.1067H,2003MNRAS.346L..57K,2013MNRAS.436.2366R}
have shown a broad distribution over a wide temperature range and a significant
amount of gas in the thermally unstable temperature range $T=500-5000\Kel$
\citep{1995ApJ...443..152W}.  Moreover, the velocity field is complex, with
turbulence dominated by large scales but potential for velocity overlap at
distinct locations along LOSs.  

Under realistic circumstances of complex velocity fields and broad temperature 
distributions, validity of the standard assumptions adopted in interpreting
\ion{H}{1} observations are open to question.  To address some of these
questions, \citet{2013MNRAS.432.3074C} have performed Monte Carlo simulations
assigning varying mass fractions of each component along each LOS, showing that
the isothermal estimator of the \ion{H}{1} column density agrees well with the
true column density within a factor of 2. However, to date there have been no
corresponding studies that construct and analyze synthetic emission/absorption
lines using results from realistic hydrodynamic simulations of the ISM.  In
this paper, we utilize our recent three dimensional hydrodynamic simulations
\citep[][hereafter Paper~I]{2013ApJ...776....1K} to create and analyze detailed
synthetic \ion{H}{1} 21~cm lines.  Our simulations enable us to investigate
fundamental questions related to use of \ion{H}{1} 21~cm lines as ISM
diagnostics. 

In addition, we can address a puzzle that has emerged from recent deep,
high-resolution interferometric observations toward radio loud quasars
\citep{2013MNRAS.436.2352R}, in combination with the LAB survey.
These observations have sufficient sensitivity to detect absorption by the WNM,
as in \citet{2003MNRAS.346L..57K}. Using emission/absorption line pairs,
\citet{2011ApJ...737L..33K} reported a threshold \ion{H}{1} column density at
$\Nth\equiv2\times10^{20}\psc$, below which the mean spin temperature exceeds
$T_s>1000\Kel$. They speculated that this apparent threshold might represent a
minimum column density for CNM to develop due to self-shielding against
ultraviolet photons. However, detailed modeling of interstellar radiation
sources have suggested that the ionization fraction remains small for the
hydrogen density of $n>10\pcc$ when the absorbing WNM column density exceeds
$10^{18}\psc$ for solar neighborhood condition (see Figure~3(c) in
\citealt{1995ApJ...443..152W}). In addition, absorption lines with a low
optical depth and a narrow line width have been detected, implying a very low
CNM column density of $\Nsim\sim5\times10^{18}\psc$
\citep{2010ApJ...722..395B}. Furthermore, \citet{2009ApJ...693.1250D} have
shown that the ratio of (integrated) emission to absorption is nearly constant
over the radial distance range of 10 to 25 kpc, whereas the emission and
absorption alone decrease by two orders of magnitude. This implies that the
mass fraction of the CNM is unchanged, while the total column density drops by
more than an order of magnitude, lower than $\Nth$. 

Here, we shall propose an alternative explanation for the \ion{H}{1} threshold
behavior observed in the solar neighborhood. This proposal is based on the
concept of the warm/cold ISM representing a self-regulated thermal and
dynamical equilibrium system that is heated and dynamically stirred by energy
injection from star formation \citep{2010ApJ...721..975O,2011ApJ...743...25K}.
In dynamical equilibrium, a given midplane pressure implies a minimum column of
gas; here, we shall show that this can give rise to a threshold column behavior
similar to that seen in observations.

The plan of this paper is as follows.  Section~\ref{sec:method} briefly reviews
the numerical models and explains how we extract LOS simulated data and produce
synthetic lines, including our procedure for spin temperature calculation. In
Section~\ref{sec:comp_sim}, we compare true values of column density, spin
temperature, and CNM mass fraction to those that would be deduced using
standard observational methods applied to our synthetic 21~cm lines.
Section~\ref{sec:comp_obs} presents mock observations for brightness
temperature, optical depth, and spin temperature for each velocity channel and
for column density distributions.  We shall show that the mock observations
reproduce \ion{H}{1} 21~cm line observations very well for individual velocity
channels \citep{2013MNRAS.436.2352R} as well as integrated over velocity
\citep{2011ApJ...737L..33K}. In addition, we show that the threshold column
density is well reproduced, and consistent with the expectation of the
thermal/dynamical model.

\section{Numerical Models and Synthetic \ion{H}{1} Line Observations}\label{sec:method}

\subsection{Brief Review of Numerical Models}\label{sec:sim}

In this paper, we utilize recent three dimensional hydrodynamic simulations of
multiphase, turbulent galactic disks presented in Paper~I to
investigate synthetic 21~cm lines. Our numerical models represent a local patch
of a galactic disk including galactic differential rotation, external gravity
from stars and dark matter, gaseous self-gravity, interstellar cooling and
heating, and star formation feedback.  We treat star formation feedback to the
neutral ISM in two ways. To represent the radiative stage of expanding
supernova remnants, we inject radial momentum, which drives turbulence at a wide
range of scale. We also apply (spatially-uniform) heating at a rate
proportional to the recent star formation rate (SFR), to model far-ultraviolet
(FUV) emission.  While simplified, this treatment incorporates both thermal and
turbulent energy feedback, both of which are required to regulate the ISM
properties and SFR properly. The reader is referred to Paper~I for complete
descriptions of model setup, numerical methods, and detailed evolution. Here,
we briefly summarize the dynamical behavior and physical properties of the disk
when it has reached a quasi-equilibrium state. 

In equilibrium, the disk models contain highly turbulent gas at a wide range of
temperatures. CNM clouds tend to settle to the midplane, where they collect to
form more massive gravitationally bound clouds (GBCs). Star formation in these
GBCs produces feedback that heats the disk and drives expanding SN shells.
Shell expansion and heating puff up the disk vertically, reducing the mean
density and rate of GBC formation. Fewer GBCs leads to lower SFR and less
feedback, until gas settles to the midplane and begins a new cycle of star
formation. After an initial transient, the models reach a state characterized
by quasi-periodic fluctuations with period\footnote{ The natural frequency of
the midplane-crossing mode is $\omega_{\rm ver}\sim\sqrt{4\pi G\rhosd}$, but
because gas streams collide at the midplane, $\tosc\sim\pi/\omega_{\rm ver}$.}
of $\tosc\sim0.5\sqrt{\pi/G\rhosd}$, where $\rhosd$ is the midplane volume
density of stars and dark matter.

After a saturated state is reached, the thermal and dynamical equilibrium model
proposed by \citet{2010ApJ...721..975O} and \citet{2011ApJ...731...41O}
describes average disk properties very well \citep[see
also][Paper~I]{2011ApJ...743...25K,2012ApJ...754....2S}. The balance between
cooling and heating sets the mean midplane thermal pressure $\Pth$ to a level
between the minimum pressure for the CNM, $\Pmin$, and the maximum pressure for
the WNM, $\Pmax$ \citep[see][]{2003ApJ...587..278W}.
\citet{2010ApJ...721..975O} assumed that $\Pth$ is approximately equal to the
geometric mean value $\Ptwo\equiv\sqrt{\Pmin\Pmax}$, which is proportional to
the heating rate. Photoelectric heating by FUV radiation, believed to be the
dominant heating process for the diffuse ISM
\citep{1995ApJ...443..152W,2003ApJ...587..278W}, is proportional to the mean
FUV radiation field and hence to the SFR surface density, $\SigSFR$.
We find the approximation $\Pth \sim \Ptwo$ holds within $60\%$ in our
numerical simulations.  

Turbulent driving and dissipation are also expected to be balanced, resulting
in the midplane turbulent pressure
$\Pturb\sim\Pdriv\equiv0.25(p_*/m_*)\SigSFR$, where $p_*$ and $m_*$ are
respectively the total radial momentum injected to the ISM by a single SN, 
and total mass in stars per SN \citep{2011ApJ...731...41O}.  Our numerical
simulations confirm that $\Pturb\sim\Pdriv$.  The total (turbulent plus
thermal) midplane pressure is then approximately linearly proportional to
$\SigSFR$.  In addition, the total midplane pressure must match the vertical
gravitational weight in order to satisfy vertical dynamical equilibrium
\citep[e.g.,][]{2007ApJ...663..183P,2009ApJ...693.1346K,2010ApJ...720.1454K,2012ApJ...750..104H}.
The level of the SFR in equilibrium is self-regulated such that all these
constraints are simultaneously satisfied. When the vertical gravitational field
is dominated by that of stars (plus dark matter), the equilibrium model
predicts $\SigSFR\propto \Pth\propto P_{\rm tot}\propto\Sigma\sqrt{\rhosd}$,
where $\Sigma$ is the gaseous surface density; our simulations show this is
satisfied \citep[][Paper~I]{2010ApJ...721..975O,2011ApJ...743...25K}.  This
situation is expected to hold for atomic-dominated regions in most galactic
disks.  

Paper~I presents results from 3D simulations at varying $\Sigma$.  Other model
parameters are set as $\rhosd\propto\Sigma^2$ and $\Omega\propto\Sigma$.  The
midplane volume density of stars and dark matter is
$\rhosd=0.05\rhounit(\Sigma/10\Surf)^2$, and the angular velocity at the center
of the local galactic patch is $\Omega=28\kms\kpc^{-1}(\Sigma/10\Surf)$.
Table~\ref{tbl} summarizes model input parameters and selected, time-averaged
physical quantities from the 3D models of Paper~I.  Column (1) gives the name
of each model from Paper~I.  Column (2) gives the gas surface density $\Sigma$.
Columns (3) to (5) give selected properties measured in the simulations
averaged over one orbital time after saturation: the SFR surface density
$\SigSFR$, the midplane thermal pressure, and the scale height of the WNM,
respectively.  We list the expected column density of the WNM-only LOS,
$\Nwnm$, in Column (6) using Equation (\ref{eq:Nwnm}).

\begin{deluxetable}{lcccccc}
\tabletypesize{\footnotesize} \tablewidth{0pt} 
\tablecaption{Model Parameters and Outcomes}
\tablehead{ 
\colhead{Model} &
\colhead{$\Sigma$} &
\colhead{$\log{\SigSFR}$} & 
\colhead{$\log{\Pth/\kbol}$} & 
\colhead{${H_w}$} &
\colhead{$\Nwnm$} \\ 
\colhead{(1)}&
\colhead{(2)}& 
\colhead{(3)}&
\colhead{(4)}& 
\colhead{(5)}&
\colhead{(6)}
}
\startdata 
QA02 &   2.5  & $-4.11\pm 0.32 $ & $2.23\pm 0.16 $ & $ 347\pm 126 $ & 0.33\\ 
QA05 &     5  & $-3.43\pm 0.27 $ & $2.70\pm 0.15 $ & $ 202\pm  50 $ & 0.56\\ 
QA10 &    10  & $-2.82\pm 0.12 $ & $3.23\pm 0.10 $ & $ 111\pm  14 $ & 1.0\\ 
QA20 &    20  & $-2.18\pm 0.06 $ & $3.79\pm 0.04 $ & $  59\pm   5 $ & 2.0
\enddata
\tablecomments{Column(1): model name from Paper~I. Column (2): gas surface
density ($\Surf$).  Column (3): logarithm of the SFR surface density
($\sfrunit$). Column (4): logarithm of the midplane thermal pressure
($\Punit$). Column (5): scale height of the WNM ($\pc$).  Column (6): fiducial
WNM-only vertical column density (see Equation (\ref{eq:Nwnm});
$10^{20}\cm^{-2}$).  Columns (3)-(5) are averaged over $t/\torb=1-2$.}
\label{tbl} 
\end{deluxetable}

\subsection{``Observing'' Simulation Output}\label{sec:los}

In order to construct synthetic 21~cm lines analogous to observations of the
Milky Way Galaxy, we assume an observer sitting at $(x, y, z)=(0, 0, 0)$, the
center of the simulation domain.  The Galactic center is located toward the
$-x$ direction and the $z=0$ plane represents the midplane of the Galactic
disk.  We use the QA10 model, a solar neighborhood analogue.  This virtual
observer conducts mock observations $10^4$ times toward random LOSs of
$(l, b)$ distributed uniformly in $dl$ and $d(\sin b)$ to ensure uniform
sampling per unit solid angle. Here, the galactic longitude $l$ is measured
counterclockwise from the $-x$ axis in the $z=0$ plane, and the galactic
latitude $b$ is measured from the midplane at $z=0$. As we briefly described in
\S\ref{sec:sim} (see also Paper~I), the model disk suffers quasi-periodic
evolutionary cycles with period $\tosc\sim 0.27\torb$. To represent the range
of disk conditions, we sample fairly in time with $\Delta t =0.1\tosc$ for a
period $2\tosc$ starting from $1\torb$.  

For each LOS, \ion{H}{1} 21~cm emission/absorption lines are constructed as
follows.  We first extract the LOS number density $n(s)$, temperature
$T_k(s)$, and velocity $v(s)$ as a function of the path length $s$ along
a pencil beam at $(l, b)$.  We hereafter omit the functional dependence on $s$
for convenience.  The local Cartesian coordinates $(x,y,z)$ can be converted to
the Galactic coordinates $(s,l,b)$ using
\begin{eqnarray}
x&=&-s\cos b \cos l \nonumber\\
y&=&-s\cos b \sin l\nonumber\\
z&=&s\sin b. \nonumber
\end{eqnarray}
Along each ray, we sample at an interval of $\Delta s=2\pc$,
calculating the physical quantities via a trilinear interpolation with the
eight nearest grid zones. The background rotational velocity relative to the
observer $\mathbf{v_0}=(0, -\Omega x, 0)$ is also added to the LOS velocity.
The path length is increased until the vertical coordinate $z$ reaches the
boundary of the simulation, $z=\pm L_z/2$, by making use of the periodic
boundary conditions in the horizontal direction.  Since our numerical model is
local, it is unphysical to perform mock observations toward very low $b$, for
which the path length would extend beyond the disk scale length
$R_d\sim3.75\kpc$ \citep{2009ARA&A..47...27K}.  For this reason, we limit the
path length to be smaller than $\smax=3\kpc$, resulting in a minimum latitude
for LOSs of $b_{\rm min}=4.9^\circ$.

\begin{figure} 
\epsscale{1.0}
\plotone{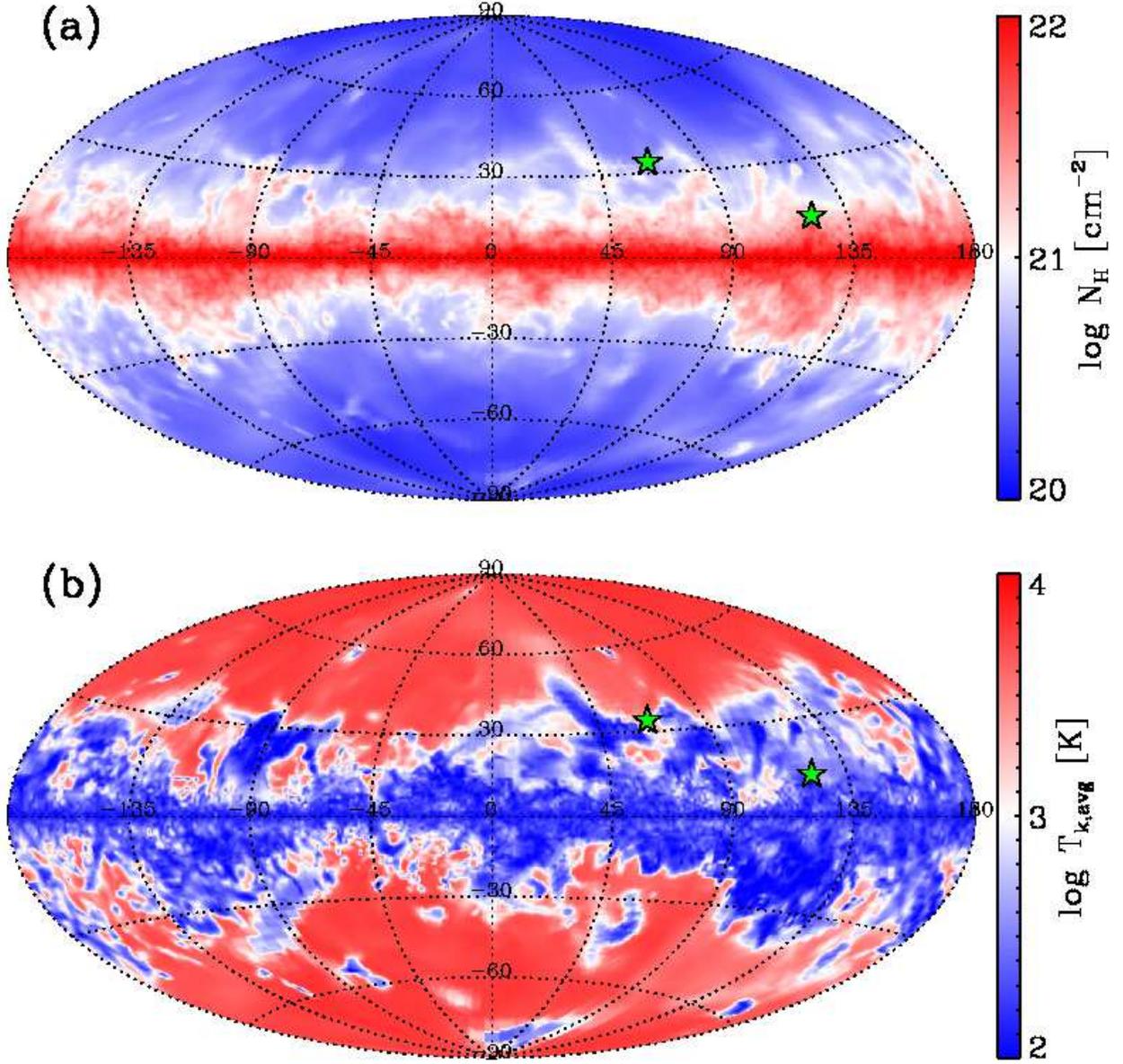}
\caption{
Aitoff projection maps of (a) column density ($\Nsim$) and (b) harmonic mean
temperature ($T_{k, {\rm avg}}$) at $t/\torb=1$. The locations of two sample
LOSs presented in Figures~\ref{fig:line_example1} and \ref{fig:line_example2}
are marked with green stars in both panels.  \label{fig:map}}
\end{figure}

In order to illustrate our virtual \ion{H}{1} sky, we calculate the column
density and the harmonic mean temperature as $\Nsim\equiv\int n ds$ and
$T_{k,{\rm avg}}\equiv\Nsim/\int (n/T_k) ds$, respectively.
Figure~\ref{fig:map} displays aitoff projection images of (a) $\Nsim$ and (b)
$T_{k,{\rm avg}}$ at $t/\torb=1$.  Here, for visualization purpose only, we
evenly divide the Galactic longitude and latitude with $1^\circ$ resolution
down to $|b|=0$, while mock observations used in this paper are randomly and
fairly sampled within an unit solid angle as described above. Note again that
the path length is artificially limited to $\smax=3\kpc$ for $|b|<5^\circ$. The
overall morphology of the column density map resembles the radio survey maps
quite well \citep[e.g.,][]{2005A&A...440..775K}.

\subsection{Synthetic \ion{H}{1} Lines}\label{sec:syn}

The emissivity and absorption coefficient\footnote{Here, the emissivity is
energy per unit volume per unit time per unit solid angle, and  we use symbol
$\kappa$ not for the opacity (the absorption cross section per mass), but for
the absorption cross section per unit volume.} of the 21~cm line are
respectively given by \citep[e.g.,][]{1978ppim.book.....S,2011piim.book.....D}
\begin{equation}\label{eq:emi}
j(\vch)=\frac{3}{16\pi}h\nu_0A_{10}n\phi(\vch)
\end{equation}
and
\begin{equation}\label{eq:abs}
\kappa(\vch)=\frac{3}{32\pi}\frac{hc^2}{k\nu_0}A_{10}\frac{n}{T_s}\phi(\vch),
\end{equation}
where $\nu_0=1.4204{\;\rm GHz}$ is the frequency at the line center,
$A_{10}=2.8843\times10^{-15}{\;\rm s}^{-1}$ is the Einstein $A$ coefficient of
the line \citep{1994ApJ...423..522G}, and $\phi(\vch)$ is the normalized line
profile as a function of velocity that satisfies $\int\phi(\vch)
\frac{\nu_0}{c}d\vch=1$.  We assume a Gaussian velocity distribution
\begin{equation}
\phi(\vch)=\frac{1}{\sqrt{\pi}\Delta v}\frac{c}{\nu_0}e^{-[(\vch-v)/\Delta v]^2},
\end{equation}
where $\Delta v\equiv\rbrackets{{2 k T_k}/{m_H}}^{1/2}$ is the Doppler width
of the line. In order to calculate the absorption coefficient, we need 
the spin temperature, $T_s$, at each point along the LOS.

The spin temperature is defined such that the Boltzmann distribution at $T_s$
gives the actual level populations.  There are three principal mechanisms that
determine the level populations of the hyperfine state: collisional
transitions, direct radiative transitions by 21~cm photons, and indirect
radiative transitions involving intermediate levels mainly by scattered
Ly-$\alpha$ photons (the Wouthuyen-Field (WF) effect;
\citealt{1952AJ.....57R..31W,1958PIRE...46..240F}).  Including all three
processes and assuming a balance between excitation and de-excitation, we can
calculate the actual level population. The spin temperature is then given by a
weighted mean of gas kinetic temperature $T_k$, the brightness temperature of
the background 21~cm radiation field $T_R=3.77\Kel$\footnote{This includes the
cosmic microwave background and Galactic synchrotron emission near the 21~cm
line \citep[][Chapter 17]{2011piim.book.....D}.}, and the effective temperature
of the Ly-$\alpha$ field $T_\alpha$ \citep[][see also
\citealt{2001A&A...371..698L}]{1958PIRE...46..240F}:
\begin{equation}\label{eq:Tspin}
T_s = \frac{T_R+y_cT_k+y_\alpha T_\alpha}{1+y_c+y_\alpha},
\end{equation}
where the normalized transition probabilities via collisions and Ly-$\alpha$
photons are respectively given by
\begin{equation}\label{eq:y}
y_c \equiv \frac{T_{0}}{T_k}\frac{R_{10}^c}{A_{10}}\quad\textrm{and}\quad
y_\alpha \equiv \frac{T_{0}}{T_\alpha}\frac{R_{10}^\alpha}{A_{10}}.
\end{equation}
Here, $R_{10}^c$ and $R_{10}^\alpha$ are the rates of net downward transition
by collision and the WF effect at given temperatures $T_k$ and $T_\alpha$,
respectively, and $T_0\equiv h\nu_0/k=0.0681\Kel$.  If collisions dominate the
transition between levels, $T_s=T_k$. However, for typical conditions in the
WNM, $T_s$ departs from $T_k$ since the rate of collisional transitions is
insufficient for ``thermalization'' \citep{2001A&A...371..698L}.

We assume that neutral hydrogen is the only collisional partner and adopt an
approximate expression for the collisional de-excitation rate coefficient
(\citealt{2011piim.book.....D}, Chapter 17; see also
\citealt{1969ApJ...158..423A,2005ApJ...622.1356Z,2006PhR...433..181F})
\begin{equation}
k_{10}\approx\left\{
\begin{array}{ll}
1.19\times10^{-10}T_2^{0.74-0.20\ln T_2} \pcc{\;\rm s}^{-1} & 20\Kel<T_k<300\Kel\\
2.24\times10^{-10}T_2^{0.207}e^{-0.876/T_2} \pcc{\;\rm s}^{-1} & 300\Kel<T_k<10^3\Kel
\end{array}
\right.,
\end{equation}
where $T_2=T_k/100\Kel$.  For $T_k>10^3\Kel$, we extrapolate the second
expression for $k_{10}$.  Collisional de-excitation due to electron impact is
less than that from neutral hydrogens provided the ionization fraction remains
less than 3\%, which is generally the case for the CNM and WNM in the solar
neighborhood \citep{1995ApJ...443..152W}. The net downward transition
rate is $R_{10}^c=nk_{10}$.

We consider the WF effect using the simple parameterized formula for $y_\alpha$
derived by \citet{1958PIRE...46..240F}:
\begin{equation}\label{eq:yalpha}
y_\alpha=5.9\times10^{11}\frac{n_{\alpha}}{T_\alpha T_k^{1/2}},
\end{equation}
where $n_{\alpha}$ is the Ly-$\alpha$ photon number density near the Ly-$\alpha$
line center.  $T_\alpha$ is determined by the spectrum near Ly-$\alpha$, which
involves a difficult scattering problem. To a first approximation,
\citet{1952AJ.....57R..31W} and \citet{1958PIRE...46..240F} argued that the
spectrum near Ly-$\alpha$ is given by the Planck curve corresponding to the
atomic kinetic temperature  so that $T_\alpha=T_k$ \citep[see
also][]{1959ApJ...129..551F}.\footnote{If there are significant non-thermal
motions of gas with one-dimensional velocity dispersion of $v_{\rm turb}$, the
Doppler temperature, $T_D\equiv T_k+m_H v_{\rm turb}^2/k$, should be considered
instead of the kinetic temperature
\citep[][]{2001A&A...371..698L,2011piim.book.....D}. Since we already rely on
specific (uncertain) parameter $n_\alpha$ to include the WF effect, however, we
limit ourselves to the simplest approximation; we note that the turbulent
velocity dispersion on large scales in our simulations does not exceed the
warm-medium thermal velocity dispersion.}  We adopt a fixed value of
$n_\alpha=10^{-6}\pcc$ for galactic Ly-$\alpha$ radiation
\citep[c.f.,][]{2001A&A...371..698L}.

Since we simply adopt a single, somewhat large value for $n_\alpha$, the
current treatment gives an upper limit of the WF effect (although the WF effect
is still insufficient to fully thermalize the WNM -- cf.
\citealt{2001A&A...371..698L}). For a lower limit on the WNM spin temperature,
we use the spin temperature neglecting the WF effect. Note that the WF effect
has little impact for the overall results in this paper in spite of about an
order of magnitude difference in the spin temperature of the WNM. This is
because we consider channel- or LOS-integrated properties, which are dominated
not by the low-optical depth WNM but by the high- or intermediate-optical depth
gas. In the remainder of this paper, we mainly present results based
calculations of $T_s$ without the WF effect (where we have found it is
unimportant), while we compare results with and without the WF effect when it
produces any non-negligible differences. 

\begin{figure*} 
\epsscale{1.0}
\plotone{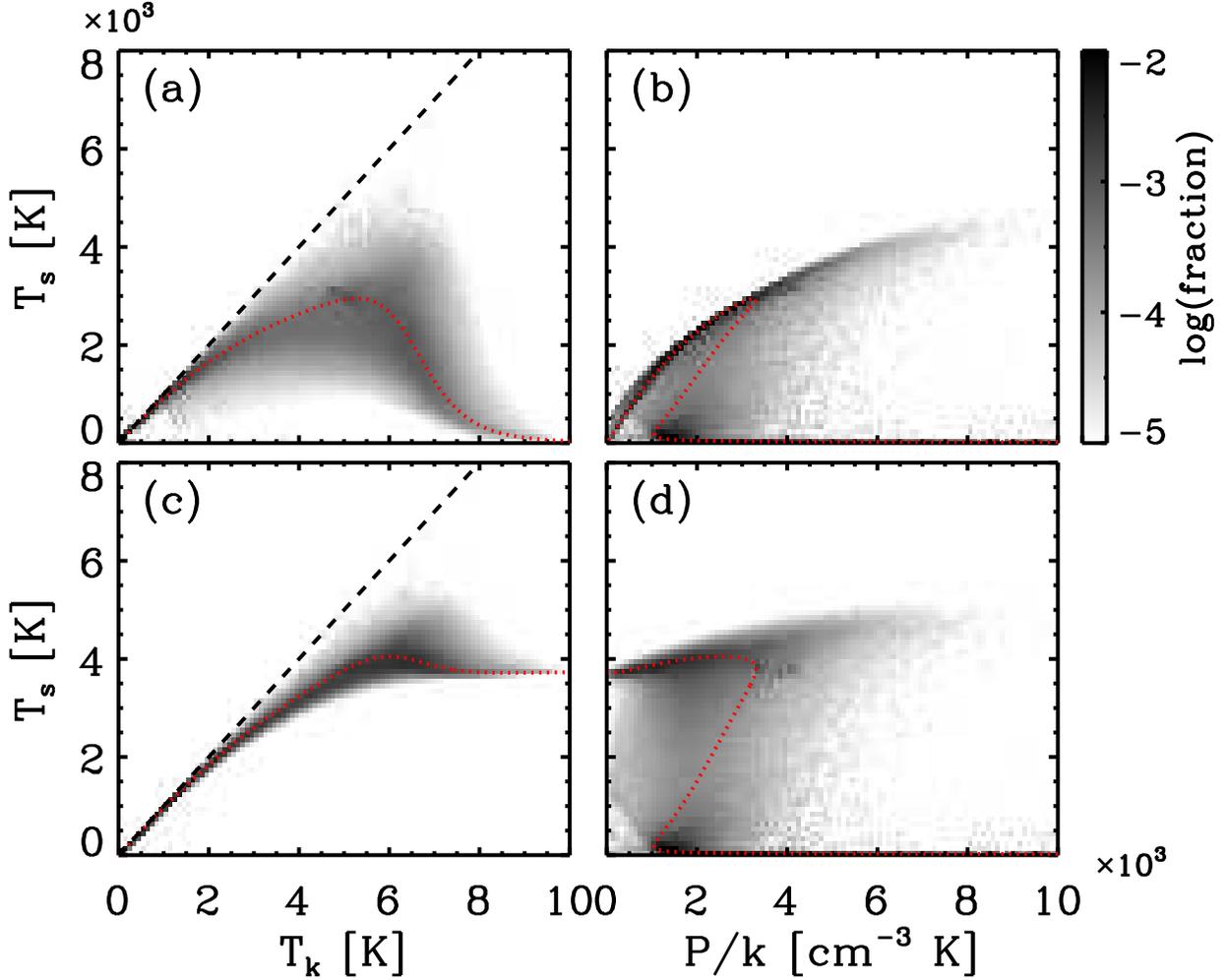}
\caption{
Distribution of the spin temperature $T_s$ without (\emph{top row}) and with
(\emph{bottom row}) the WF effect as a function of the kinetic temperature
$T_k$ (\emph{left column}) the thermal pressure $P/k$ (\emph{right column}) for
model QA10 at $t/\torb=1$.  The grey scale displays the logarithmic fraction of
the total mass for all simulated LOSs in each bin.  The black dashed line in
(a) and (c) stands for $T_s=T_k$.  The red dotted line denotes the spin
temperature taking density equal to the thermal equilibrium value, $n_{\rm
eq}=\Gamma/\Lambda(T_k)$, at a given $T_k$.  At $T_k>5000\Kel$, the equilibrium
spin temperature drops (strongly without WF in (a), slightly with WF in (c)).
Panels (b) and (d) show that this corresponds to low-pressure gas with
$P/k<10^3\pcc\Kel$, away from the midplane in the simulation.
Out-of-equilibrium gas is also evident for $T_k>1000\Kel$.  } \label{fig:Tspin}
\end{figure*}

Figure~\ref{fig:Tspin} displays the distribution of the spin temperature $T_s$
without the WF effect (\emph{top row}; (a) and (b)) and with the WF effect
(\emph{bottom row}; (c) and (d)) as a function of the kinetic temperature $T_k$
(\emph{left column}; (a) and (c)) and the thermal pressure $P/k = 1.1 n T_k$
(\emph{right column}; (b) and (d)) for the snapshot at $t/\torb=1$ of the QA10
model.  The grey scale represents the fraction of the total mass in each bin.
The red dotted line denotes the value of the spin temperature calculated along
the thermal equilibrium density curve, $n_{\rm eq}\equiv\Gamma/\Lambda(T_k)$,
for the average heating rate of the QA10 model $\Gamma=1.6\times10^{-26}\ergs$
and using $\Lambda$ given by Equation (6) of Paper~I.  

Figure~\ref{fig:Tspin} shows that $T_s\sim T_k$ for the gas at $T_k<500\Kel$,
whereas a range of $T_s$ is possible for the warm and unstable gas with
$T_k>1000\Kel$.  Note however that low values of $T_s$ in warm gas (without the
WF effect) occur only when the pressure is quite low, far from the midplane.
The WF effect brings $T_s$ closer to $T_k$ at low density and pressure (panels
(c) and (d)), but the transition is still not thermalized.  Since most of the
gas is close to thermal equilibrium (because the cooling time is shorter than
the dynamical time), the most populated part of the distribution follows the
equilibrium curve.  However, a non-negligible fraction of gas is
out-of-equilibrium with higher/lower density than $n_{\rm eq}$ due to dynamical
contraction/expansion, resulting in higher/lower spin temperature than the red
dotted line (see also Figures~8 and 10 of Paper~I).  In practice, without the
WF effect, the density range for out-of-equilibrium gas is quite wide (more
than an order of magnitude) so that $T_s$ spans from a few hundreds of Kelvin
to $T_k$ when $T_k>1000\Kel$.  With the WF effect, $T_s$ of the WNM is
generally larger than $1000\Kel$ but still significantly smaller than $T_k$.

Given $T_s$, we now can calculate the absorption coefficient from Equation
(\ref{eq:abs}), and the contribution to the local optical depth from the $i$-th
element along a LOS is then $\tau^i(\vch)=\kappa(\vch;s_i)\Delta s$, where
$s_i=i\Delta s$. The total optical depth along a LOS for a given velocity
channel $\vch$ is then:
\begin{equation}\label{eq:tau}
\tau(\vch)=\sum_{i=1}^N \tau^i(\vch),
\end{equation}
where $N$ is the total number of LOS elements in a given $(l,b)$ direction.
The synthetic emission line is constructed by summing up the brightness
temperature of each element with foreground attenuation:
\begin{equation}\label{eq:TB}
T_B(\vch)=\sum_{i=1}^N\sbrackets{ T_s^i
(1-e^{-\tau^i(\vch)})\exp\rbrackets{-\sum_{j=1}^{i-1}\tau^j(\vch)}}.
\end{equation}
We use velocity channels with resolution $\Delta \vch = 1\kms$ from $-50\kms$
to $50\kms$.

\begin{figure} 
\epsscale{0.8}
\plotone{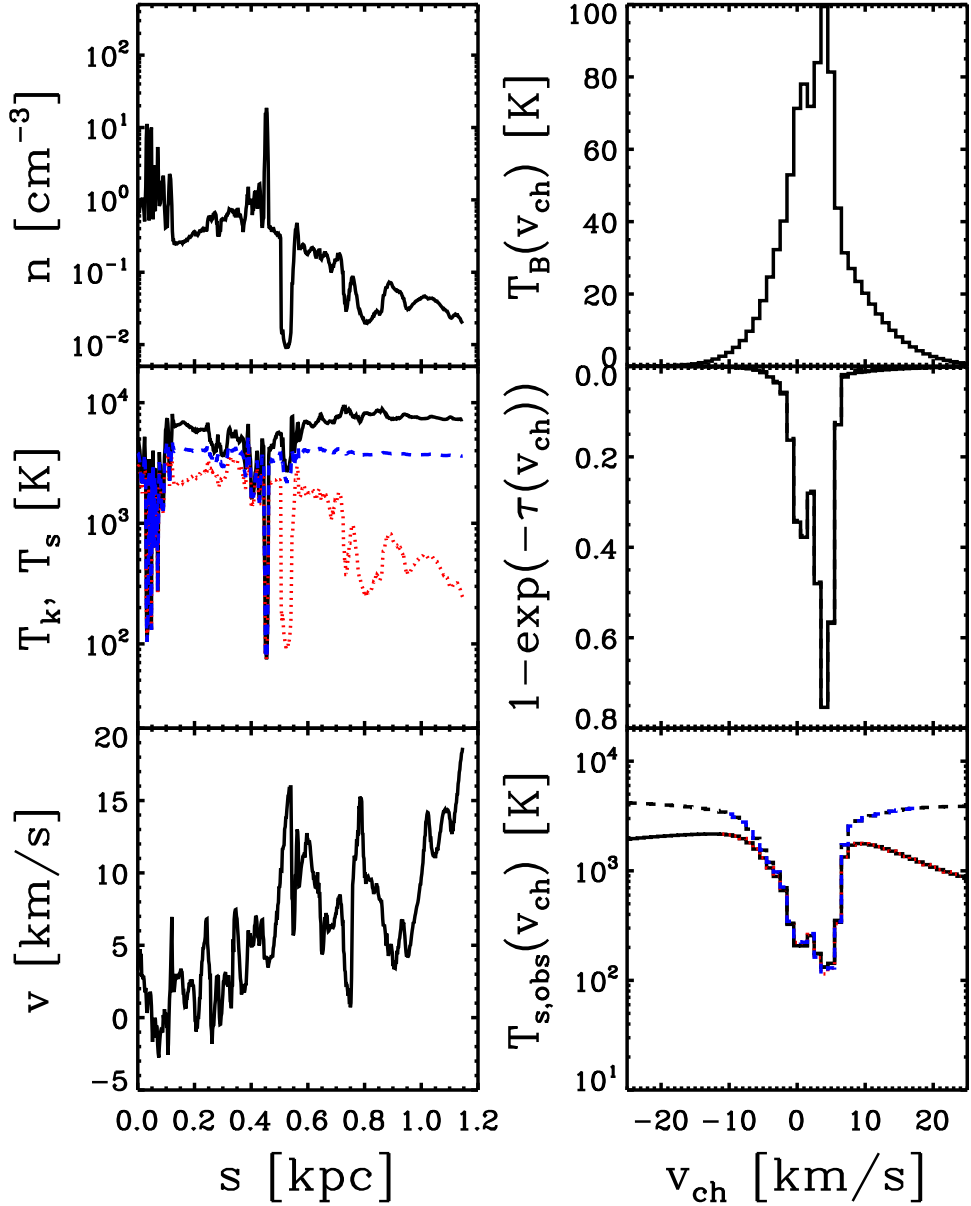}
\caption{
Example LOS toward direction with high-$\tau$, $(l, b)=(121^\circ, 12^\circ)$
at $t/\torb=1$.  \emph{Left}: From \emph{top} to \emph{bottom}, we plot in
black the number density $n$, temperature $T_k$, and velocity $v$ as functions
of the path length $s$ along the LOS.  In the \emph{middle} plot, the spin
temperature $T_s$ is shown with (blue dashed) and without (red dotted) the WF
effect.  Without WF, $T_s$ drops at large distance from the midplane where the
pressure is low.  \emph{Right}: From \emph{top} to \emph{bottom}, we plot the
brightness temperature $T_B(\vch)$ (synthetic emission line), the optical depth
$1-e^{-\tau(\vch)}$ (synthetic absorption line), and the ``observed'' spin
temperature $\Tsobs\equiv T_B(\vch)/(1-e^{-\tau(\vch)})$ as functions of
velocity channel.  In the \emph{middle} and \emph{bottom} plots, solid black
lines are without and dashed lines are with the WF effect.  
Note that solid and dashed lines in the \emph{middle} plot are
indistinguishable for high-$\tau$ channels (see Figure~\ref{fig:line_example2}
for low-$\tau$ channels).
In the \emph{bottom} plot, the harmonic mean spin temperature $\Tsavg$ defined
by Equation (\ref{eq:Tsavg}) is show with (blue dashed) and without (red
dotted) the WF effect.  The ``observed'' spin temperature agrees with the
``true'' harmonic mean spin temperature either with or without the WF effect.
See Section~\ref{sec:NHTS} for details.}\label{fig:line_example1} 
\end{figure}

\begin{figure} 
\epsscale{0.8}
\plotone{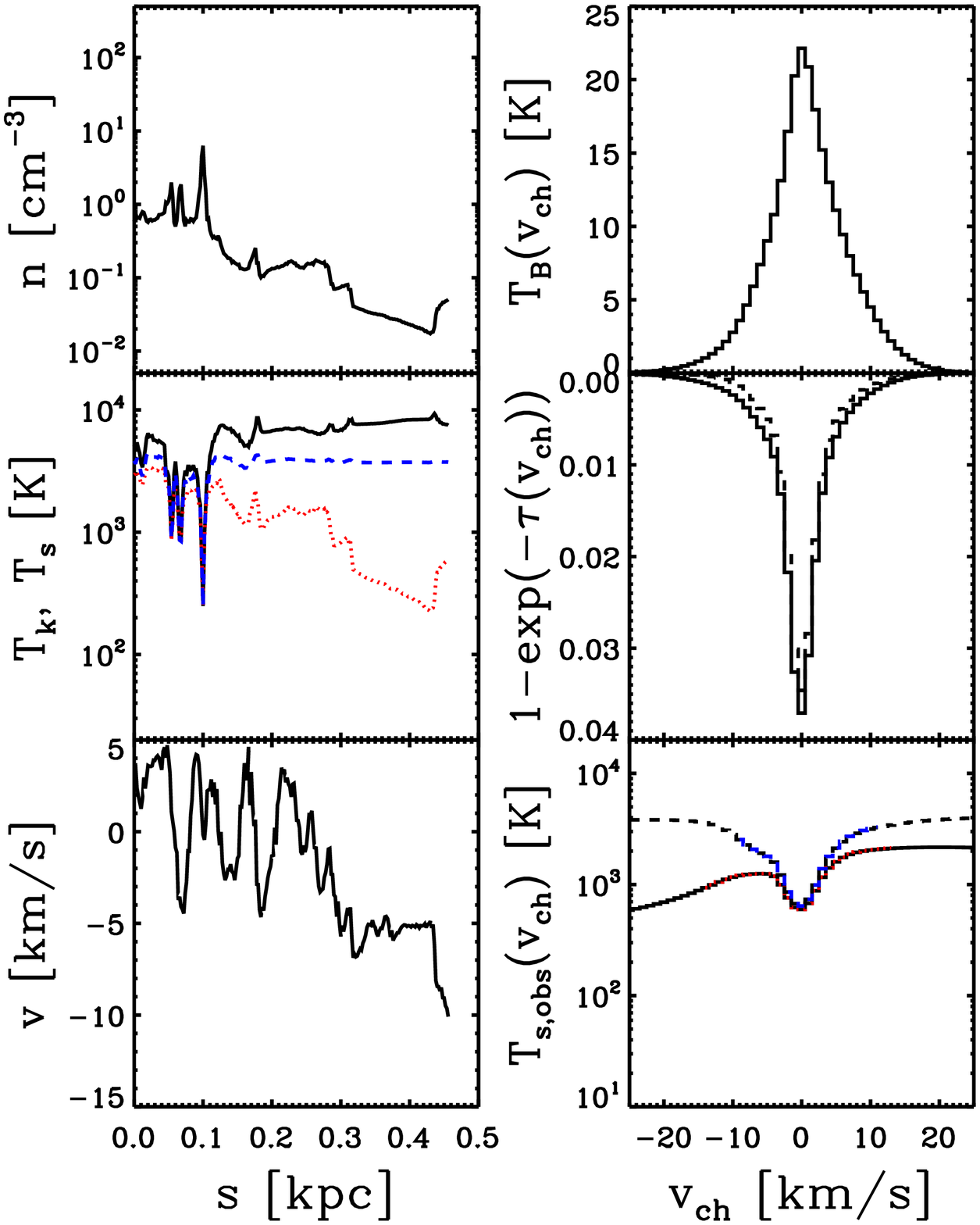}
\caption{
Same as Figure~\ref{fig:line_example1} but for low-$\tau$ LOS, $(l, b)=(66^\circ, 33^\circ)$. 
}\label{fig:line_example2} 
\end{figure}

Figures~\ref{fig:line_example1} and \ref{fig:line_example2} plot examples of
high and low optical depth LOSs toward $(l, b)=(121^\circ, 12^\circ)$ and $(l,
b)=(66^\circ, 33^\circ)$ at $t/\torb=1$, respectively.  In the \emph{left}
column, we plot (\emph{top} to \emph{bottom}) the number density, temperature,
and velocity versus path length along the LOS.  In the \emph{right} column, we
plot the synthetic emission ($T_B(\vch)$) and absorption ($1-e^{-\tau(\vch)}$)
lines as well as the ``observed'' spin temperature $\Tsobs$ (see Equation
(\ref{eq:Tsobs})) as a function of velocity channel.  The spin temperature
$T_s$ without the WF effect at each point along the LOS, and the harmonic mean
spin temperature in each velocity channel, $\Tsavg$ (see Equation
(\ref{eq:Tsavg})), are presented as red dotted lines in \emph{middle-left} and
\emph{bottom-right} plots, respectively, while the blue dashed lines denote
profiles with the WF effect.  Note that the optical depth is slightly decreased
due to the WF effect (dashed line in \emph{middle-right} plot), while the
synthetic emission remains unchanged.  We only plot $\Tsavg$ for
$\tau(\vch)>10^{-3}$ to show the range of synthetic lines we use for further
analysis.  From Figure~\ref{fig:line_example1}, it is evident that density
structure is quite complicated along a given LOS, even when $T_B(\vch)$ and
$\tau(\vch)$ are relatively simple.

\section{Extraction of Physical Properties from Synthetic Lines}\label{sec:comp_sim}

\subsection{Column Density and Spin Temperature}\label{sec:NHTS}

The column density and harmonic mean spin temperature are the main physical
properties that can be inferred directly from the \ion{H}{1} 21~cm emission and
absorption lines observations
\citep[e.g.,][]{2003ApJ...586.1067H,2005A&A...440..775K,2009ApJ...693.1250D}.
From Equations (\ref{eq:abs}) and (\ref{eq:tau}), integrating the density along
the LOS gives
\begin{equation}\label{eq:NHsim}
\Nsim =\int n ds= 1.813\times10^{18}\cm^{-2}\int \Tsavg \tau(\vch) d\vch
\quad(d\vch \textrm{ in} \kms), 
\end{equation}
where the harmonic mean spin temperature for a given velocity channel is defined by
\begin{equation}\label{eq:Tsavg}
\Tsavg\equiv\frac{\int\kappa(\vch)T_s ds}{\tau(\vch)}= 
\frac{\int \kappa(\vch) T_s ds}{\int \kappa(\vch) ds}=
\frac{\int n\phi(\vch) ds}{\int(n/T_s)\phi(\vch) ds}.
\end{equation}
Note, however, that $\Tsavg$ is not a direct observable since the definition
in Equation (\ref{eq:Tsavg}) requires full LOS information. For observations, an
``observed'' spin temperature in a given velocity channel is defined by
\begin{equation}\label{eq:Tsobs}
\Tsobs\equiv\frac{T_B(\vch)}{1-e^{-\tau(\vch)}},
\end{equation}
where $T_B(\vch)$ is the observed brightness temperature, and
$1-e^{-\tau(\vch)}$ is obtained by differencing off and on a background source.
By inspecting Equation (\ref{eq:TB}), the approximation $\Tsobs\approx\Tsavg$
holds for channels with low optical depth and absorption dominated by a single
layer \citep{1978ppim.book.....S}.  By comparing $\Tsavg$ (red and blue) and
$\Tsobs$ (black solid and dashed) in \emph{bottom-right} panels of
Figures~\ref{fig:line_example1} and \ref{fig:line_example2}, we can see that
the ``observed'' spin temperatures inferred from the lines in fact agree quite
well with the corresponding harmonic mean spin temperatures even near the line
center with large optical depth.  This is true whether or not WF effect
contributes to setting the spin temperature.

\begin{figure} 
\epsscale{1.0}
\plotone{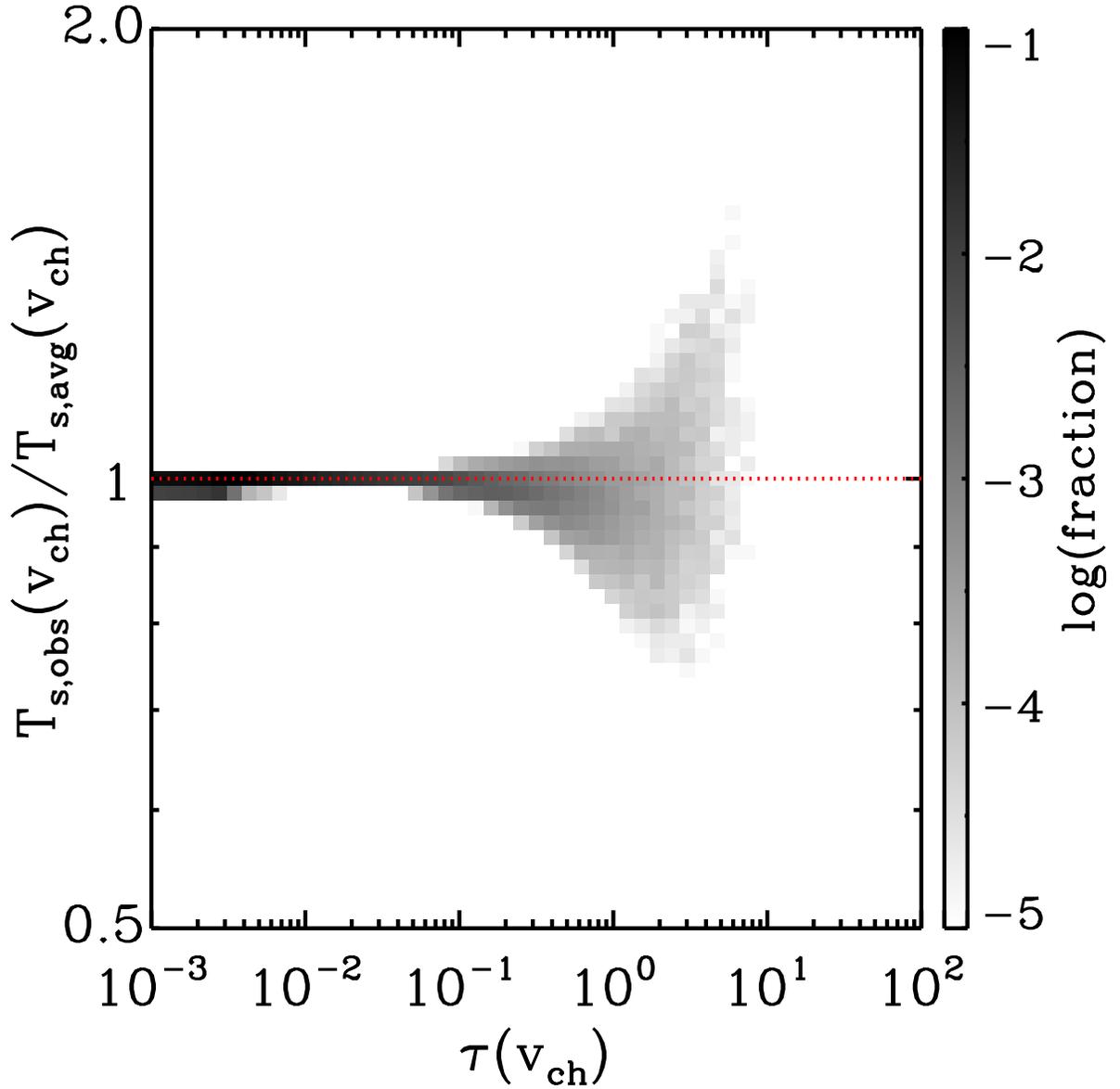}
\caption{
Distribution of the ratio of the ``observed'' to ``true'' spin temperatures
$\Tsobs/\Tsavg$ as a function of the channel optical depth $\tau(\vch)$ for all
LOSs and channels at $t/\torb=1$.  The grey scale displays the fraction of the
whole distribution in each bin.  With the WF effect, this distribution remains
unchanged.  }\label{fig:Tscomp} 
\end{figure}

In order to quantify the agreement between the ``observed'' and ``true''
harmonic-mean spin temperatures, we calculate $\Tsobs/\Tsavg$ as a function of
the channel optical depth $\tau(\vch)$ for all LOSs and channels of a snapshot
at $t/\torb=1$.  Figure~\ref{fig:Tscomp} displays the distribution for the case
without WF in grey scale.  As expected, if the channel is optically thin
$\tau(\vch)\simlt0.1$, the two spin temperatures are in very good agreement.
The difference emerges at $\tau(\vch)\simgt 0.1$ and gets larger as the channel
optical depth increases.  $\Tsobs$ can either under- or overestimate the
``true'' harmonic mean spin temperature, but remains within a factor of 1.2 and
1.5 of $\Tsavg$ for the channel optical depth of $\tau(\vch)\sim1$ and $10$,
respectively.  Note that $T_s$ is larger when the WF effect is included, which
reduces both the contribution to the harmonic-mean average spin temperature and
the contribution to the optical depth, from warm regions in each velocity
channel.  The agreement between $\Tsobs$ and $\Tsavg$ is best at low optical
depth, so the use of lower limit for the WNM spin temperature (omitting WF)
provides a conservative conclusion for the agreement between ``observed'' and
``true'' values.  We have also directly checked that when the WF effect is
included, Figure~\ref{fig:Tscomp} is essentially unchanged.  

We now explore using $\Tsobs$ as a proxy for $\Tsavg$ to obtain the
``observed'' column density from the synthetic lines.  Substituting Equation
(\ref{eq:Tsobs}) for $\Tsavg$ in Equation (\ref{eq:NHsim}),
\begin{equation}\label{eq:Nsyn}
\Nsyn\equiv 1.813\times10^{18}\cm^{-2}
\int\frac{\tau(\vch)T_B(\vch)}{1-e^{-\tau(\vch)}}d \vch
 \quad(d\vch \textrm{ in} \kms). 
\end{equation}
The quantities on the right-hand side are all observables; Equation
(\ref{eq:Nsyn}) is also known as the ``isothermal'' estimator of the \ion{H}{1}
column density
\citep[][]{1978ppim.book.....S,1982AJ.....87..278D,2013MNRAS.432.3074C}. In the
optically-thin limit, we have the ``thin'' column density
\begin{equation}\label{eq:Nthin}
\Nthin\equiv 1.813\times10^{18}\psc\int T_B(\vch) d \vch
 \quad(d\vch \textrm{ in} \kms),
\end{equation}
which is what observers obtain when there is no absorption line information.
The ``observed'' spin temperature obtained from an optical-depth weighted
average in a given LOS is:
\begin{equation}\label{eq:Tsyn}
\Tsyn\equiv \frac{\int\tau(\vch)\Tsobs d\vch}{\int \tau(\vch)d \vch}.
\end{equation}

\begin{figure*} 
\epsscale{1.0}
\plotone{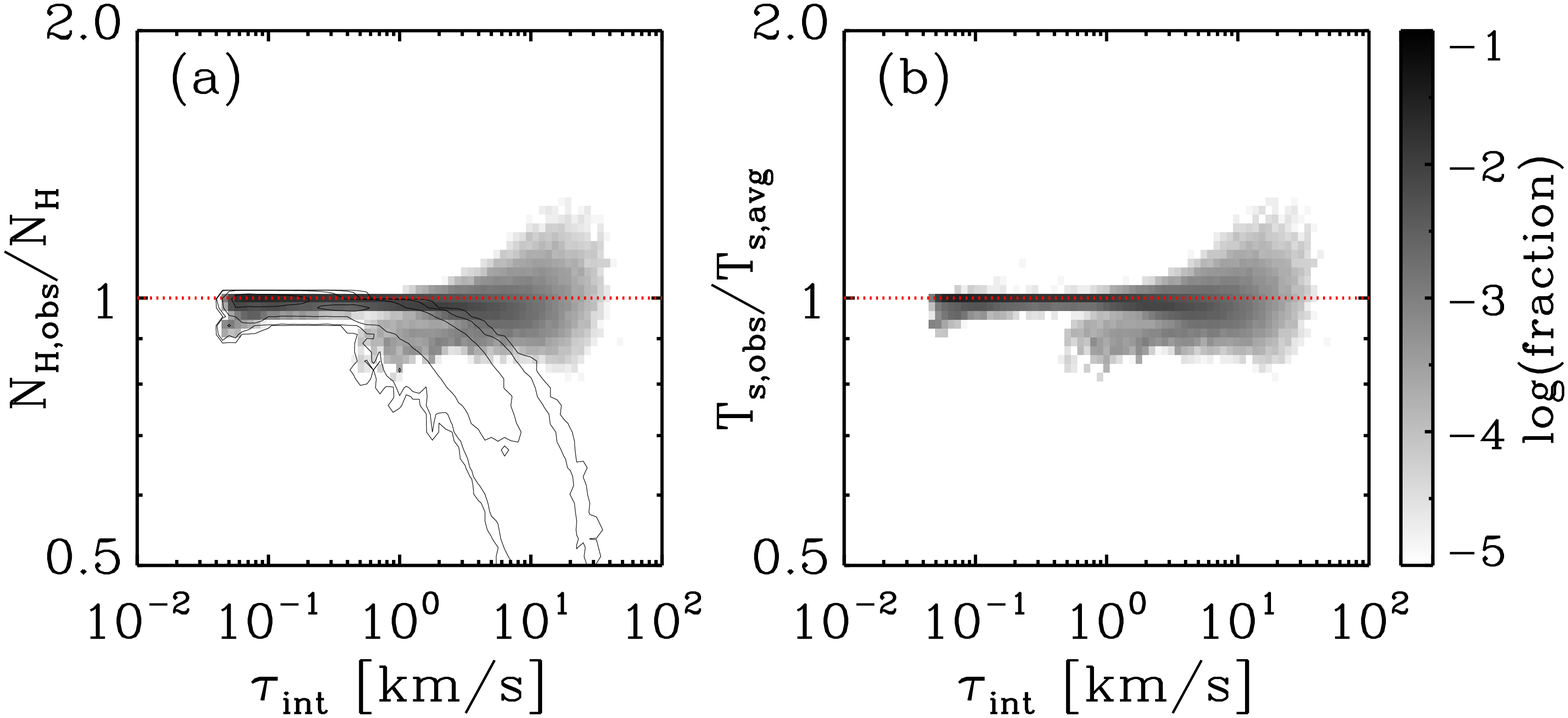}
\caption{
Distribution (grey scale) of the mock observation data of the ratios of
``observed'' to ``true'' (a) column density $\Nsyn/\Nsim$ and (b) harmonic mean
spin temperature $\Tsyn/\Tsim$ as a function of the integrated optical depth
$\tauint\equiv\int\tau(\vch)d\vch$ for all LOSs and simulation snapshots.  The
thin contours show the distribution for the ratio of ``thin'' to ``true''
column density $\Nthin/\Nsim$.  The contour levels from outside to inside
correspond to the number fractions of $10^{-5}$, $10^{-4}$, $10^{-3}$,
$10^{-2}$.  See Equations (\ref{eq:Nsyn}), (\ref{eq:Nthin}), and
(\ref{eq:Tsyn}) for definitions of $\Nsyn$, $\Nthin$, and $\Tsyn$,
respectively.}\label{fig:NHcomp} 
\end{figure*}

Figure~\ref{fig:NHcomp} displays distributions of the mock observational data
for the ratios of ``observed'' to ``true'' (a) column density $\Nsyn/\Nsim$ and
(b) harmonic mean spin temperature $\Tsyn/\Tsim$ as a function of the
integrated optical depth $\tauint\equiv\int\tau(\vch)d\vch$.
Figure~\ref{fig:NHcomp} includes all LOSs and all temporal snapshots over an
interval $2\tosc$ (rather than a single data dump) from the simulation.  Here,
the ``true'' harmonic mean spin temperature is defined by $\Tsim\equiv\int n
ds/\int (n/T_s) ds$, which can also be obtained by taking the optical depth
weighted average of $\Tsavg$ analogous to Equation (\ref{eq:Tsyn}).  Since
$\Tsobs$ reproduces $\Tsavg$ very well, $\Nsyn$ and $\Tsyn$ also give excellent
estimates of the true column density and harmonic mean spin temperature,
respectively.  In particular, 90\% and 99\% of all LOSs respectively have
$\Nsyn$ and $\Tsyn$ within 5\% and 12\% of the true values.  As in
Figure~\ref{fig:NHcomp}, the estimators are best at low $\tauint$.  The bulk of
underestimated data points shown near $\tauint\sim 0.5$ are due to a specific
snapshot at $t/\torb=1.3$, when a cold cloud happens to surround the virtual
observer. The cold cloud surrounding the observer provides foreground
absorption at moderate opacity for all LOSs, violating the assumption of a
single cold layer and resulting in overall underestimation of $\Tsobs$ and
hence $\Nsyn$ and $\Tsyn$. 

The distribution of the ratio of the ``thin'' (Equation \ref{eq:Nthin}) to
``true'' column density is shown as thin contours in
Figure~\ref{fig:NHcomp}(a).  Evidently, $\Nthin$ significantly underestimates
the ``true'' column density when $\tauint>1$. However, the majority of the
$\Nthin/\Nsim$ distribution remains $>0.7$ up to $\tauint<10$, while the
distribution extends to values as small as $\sim 0.2$ at high $\tauint$. In
observations, $\Nthin/\Nsyn\sim0.6-0.8$ for $\Nsyn\sim4-20\times10^{21}\psc$
\citep[][see also \citealt{1982AJ.....87..278D}]{2003ApJ...585..801D}.

\subsection{CNM MASS FRACTION}\label{sec:fcnm}

Another important physical property that can be deduced from
emission/absorption line observations is the CNM mass fraction
\citep[e.g.,][]{2009ApJ...693.1250D}. In the classical two phase picture of the
ISM \citep[e.g.,][]{1969ApJ...155L.149F,1995ApJ...443..152W}, the harmonic mean
kinetic temperature $T_{k, {\rm avg}}$ along the LOS is defined by
\begin{equation}\label{eq:fcnm}
\frac{\Nsim}{T_{k, {\rm avg}}}\equiv\frac{N_c}{T_c}+\frac{N_w}{T_w},
\end{equation}
where $N_c$ and $N_w$ are the column density of the CNM and WNM, respectively,
and $T_c$ and $T_w$ are the temperature of the CNM and WNM, respectively.
Since $T_c\ll T_w$, the CNM mass fraction $\fcsim\equiv N_c/\Nsim\approx
T_c/T_{k, {\rm avg}}$ unless $\fcsim$ is extremely small. 

Similarly, for a two phase ISM, Equation (\ref{eq:Tsyn}) can be simplified
(using Equation (\ref{eq:NHsim}) and $\Tsobs\approx\Tsavg$) as
\begin{equation}\label{eq:Tsyn2}
\frac{\Nsyn}{\Tsyn}\approx\frac{N_c}{T_{s,c}}+\frac{N_w}{T_{s,w}}
\approx\frac{N_c}{T_c}+\frac{N_w}{T_{s,w}},
\end{equation}
where $T_{s,c}$ and $T_{s,w}$ are the spin temperatures of the CNM and
WNM, respectively.  The second approximation in Equation (\ref{eq:Tsyn2})
utilizes $T_{s,c} \approx T_c$ from Figure~\ref{fig:Tspin}. This gives
\begin{equation}\label{eq:fcsyn}
\fcsyn\equiv\frac{N_c}{\Nsyn}=\frac{T_c}{\Tsyn}\rbrackets{\frac{T_{s,w}-\Tsyn}{T_{s,w}-T_c}}.
\end{equation}
Here, we have used $\Nsyn\approx \Nsim$ and we keep the term in parentheses
since typical spin temperatures of the WNM are not as high as the kinetic
temperature of the WNM (see Figure~\ref{fig:Tspin}). Thus, for a given $\Tsyn$,
one can estimate the CNM mass fraction by assuming a value of $T_c$ and
$T_{s,w}$. If $T_{s,w}\gg\Tsyn$, then $\fcsyn\sim T_c/\Tsyn$.

\begin{figure} 
\epsscale{0.8}
\plotone{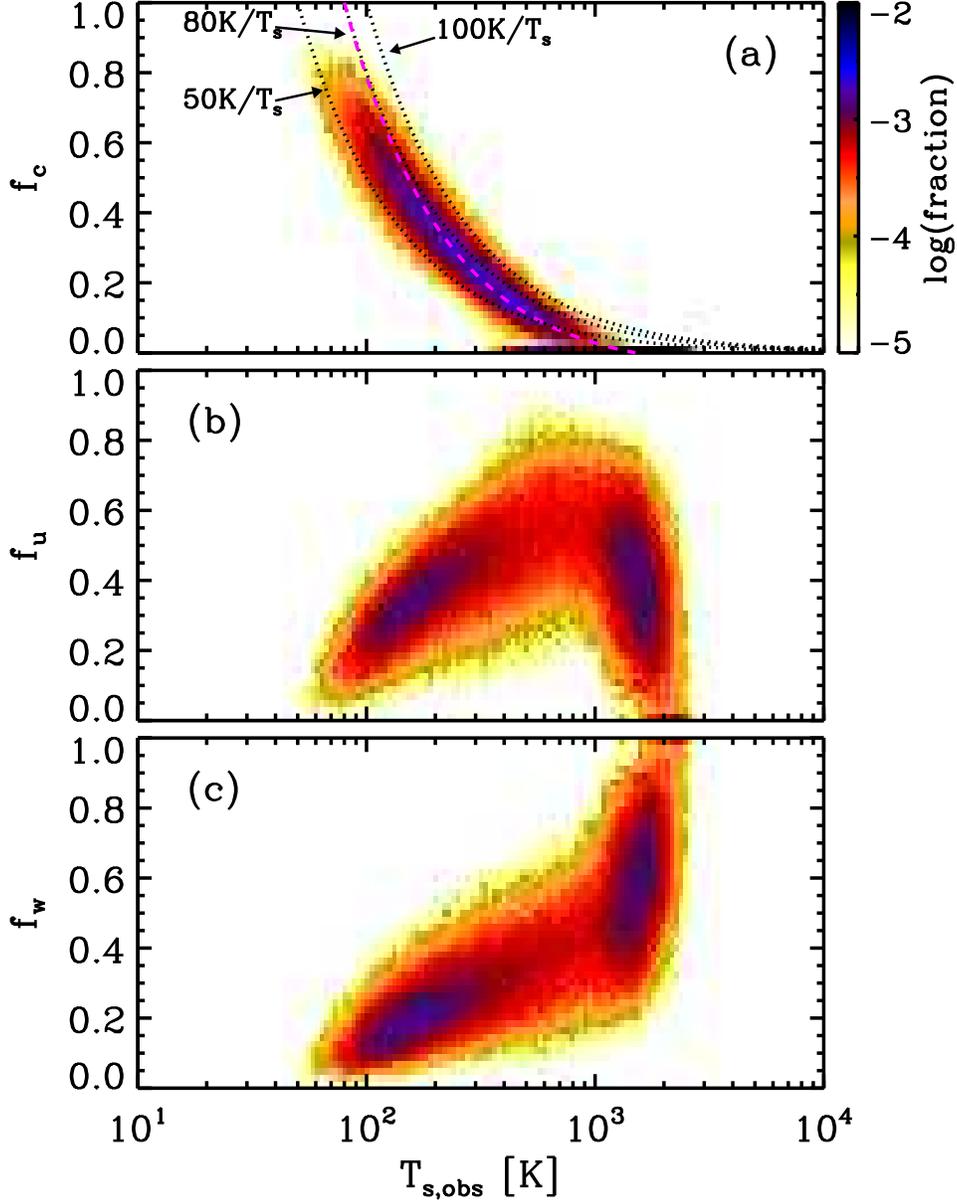}
\caption{
Distribution of the  mass fractions of (a) cold ($f_c$; $T_k<184\Kel$),
(b) thermally unstable ($f_u$; $184\Kel<T_k<5050\Kel$), and (c) warm ($f_w$;
$T_k>5050\Kel$) gas as a function of the ``observed'' mean spin temperature for
all LOS and simulation snapshots. The black dotted lines in (a) denote
$\fcsyn\approx T_c/\Tsyn$ for $T_c=50\Kel$, $80\Kel$, and $100\Kel$ from bottom
to top. For $\Tsyn<400\Kel$, the $T_c=80\Kel$ curve agrees well with $f_c$, and
the $T_c=50\Kel$ and $T_c=100\Kel$ lines envelope the overall distribution. The
magenta dashed line shows $\fcsyn$ (see Equation (\ref{eq:fcsyn})) with $T_{\rm
s,w}=1500\Kel$; this follows densest part of the distribution very well. 
\label{fig:fcnm}}
\end{figure}

In order to test the feasibility of this method, we first calculate the
``true'' mass fractions of the CNM ($f_c$; $T_k<184\Kel$), unstable neutral
medium (UNM, $f_u$; $184\Kel<T_k<5050\Kel$), and WNM ($f_w$; $T_k>5050\Kel$)
for a given LOS by integrating number density $n$ of each component directly.
Figure~\ref{fig:fcnm} plots the distribution of the ``true'' mass fractions of
(a) CNM (b) UNM and (c) WNM as a function of the ``observed'' spin temperature
$\Tsyn$ without the WF effect for all LOSs and snapshots.  For $\Tsyn<400\Kel$,
the main distribution of $f_c$ follows roughly $\fcsyn\approx80\Kel/\Tsyn$, the
``observed'' CNM mass fraction.  Here, we adopt $T_c=80\Kel$ based on the mean
temperature of the CNM in the simulation.  The scatter of the distribution
indicates that the temperature of the CNM is not a constant but spans a range
of $50\Kel<T_c<100\Kel$ (see the dotted lines in Figure~\ref{fig:fcnm}(a)).
For $\fcsim<0.2$ with $\Tsyn>400\Kel$, however, this simple approximation for
$f_c$ employing $T_c$ alone is no longer valid. Instead, by adopting
$T_{s,w}=1500\Kel$ (the typical spin temperature of the WNM without the WF
effect -- see Figure~\ref{fig:Tspin}(a) and (b)), we find that $\fcsyn$ from
Equation (\ref{eq:fcsyn}) follows the overall trend of the $\fcsim$
distribution quite well (see the dashed line in Figure~\ref{fig:fcnm}(a)). In
any case, the intrinsic scatter of this mass fraction estimator is as high as a
factor of 2.

Figure~\ref{fig:fcnm}(b) and (c) show that there is no regime in which either
the UNM or the WNM is individually predominant in any range of the spin
temperature.  For the range of $200\Kel<\Tsyn<1000\Kel$, the UNM is in the
majority, but the WNM mass fraction is nearly comparable. For the range of
$\Tsyn>1000\Kel$, the WNM becomes the majority, but the UNM fraction is still
not negligible. The most probable UNM mass fraction is 30-50\% for
$\Tsyn>1000\Kel$.  This implies that any attempt to use the spin temperature to
estimate relative UNM and WNM abundances would not be reliable. Observation of
spin temperature in the regime $\Tsyn\simgt1000\Kel$ only implies that there is
negligible CNM.

\begin{figure} 
\epsscale{0.8}
\plotone{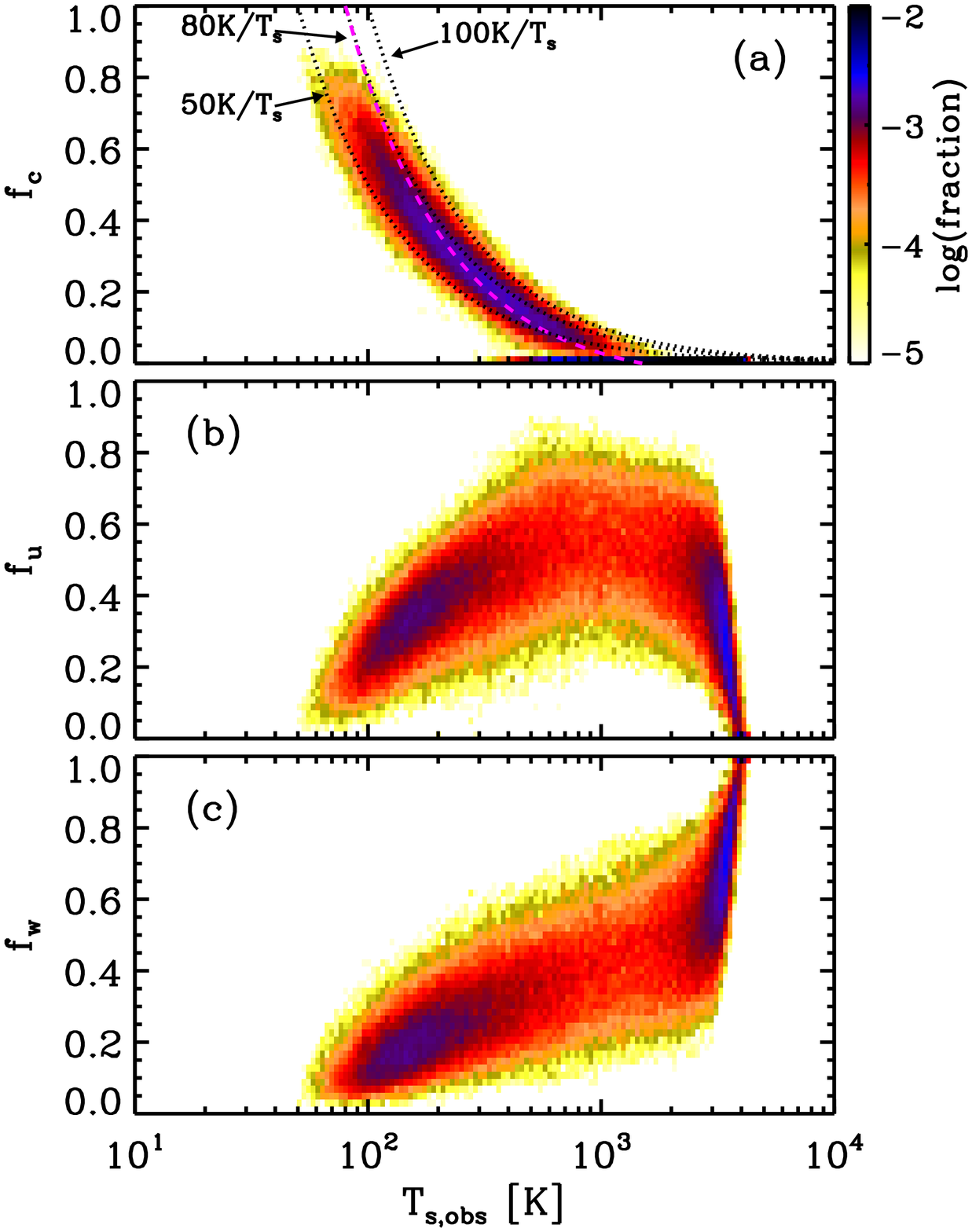}
\caption{
Same as Figure~\ref{fig:fcnm} but with the WF effect. Since $T_s$ is larger
with the WF effect for $T_k>1000\Kel$ and distributed in a narrower temperature
range (see Figure~\ref{fig:Tspin}), the WNM dominated region moves toward
higher $\Tsyn$ with a sharp transition at around $\Tsyn=4000\Kel$.  Note that
the distribution of $f_c$ remains unchanged by the WF effect, and is well
described by $\fcsyn$ with $T_{\rm s,w}=1500\Kel$ (magenta dashed).
\label{fig:fcnm_WF}}
\end{figure}

Figure~\ref{fig:fcnm_WF} shows the same distributions of phases as a function
of the ``observed'' spin temperature $\Tsyn$, but now with the WF effect
included.  The distributions of all gas components are unchanged at
$\Tsyn<1000\Kel$, while $f_u$ and $f_w$ distributions are stretched toward
higher $\Tsyn$ at $\Tsyn>1000\Kel$.  When the WF effect is included, $T_s$
spans a narrower range at $\Tsyn>1000\Kel$ (see Figure~\ref{fig:Tspin}), and as
a result $f_w \rightarrow 1$ sharply at around $\Tsyn\sim4000\Kel$.  Note
however that the value of $T_s$ above $1000\Kel$ is quite uncertain, as it
depends on poorly constrained parameters controlling the WF effect; the
specific values of $f_w$ shown here in that range should therefore be
considered cautiously.  Taken together, Figures~\ref{fig:fcnm} ($T_s$ without
WF) and \ref{fig:fcnm_WF} ($T_s$ with WF) imply that the typical ``observed''
spin temperature of the WNM (including substantial amount of the UNM) would be
in a range between $1000$-$5000\Kel$, consistent with previous calculation by
\citet{2001A&A...371..698L}. 

\section{Comparison with Observations}\label{sec:comp_obs}

\subsection{Brightness Temperature, Optical Depth, and Spin Temperature}

Classically, \ion{H}{1} emission/absorption line observations have reported a
negative correlation between  the peak optical depth $\taupeak$ and the spin
temperature at the peak $\Tspeak$ since \citet{1975A&A....42...25L} first noted
this correlation. The typical slope between $\log \Tspeak$ and $\log \taupeak$
is found to be $-0.35$
\citep[e.g.,][]{1983ApJ...272..540P,1987ASSL..134...87K}.  As noted by
\citet{2003ApJ...586.1067H} (see also \citealt{1992ApJ...386..120B}), however,
not only $\Tspeak$ and $\taupeak$ but also $\Nsim$ and $\vfwhm$ are mutually
related as
\begin{equation}\label{eq:Ttau}
\Nsim=1.93\times10^{20}\psc\;\taupeak\rbrackets{\frac{\Tspeak}{100\Kel}}
\rbrackets{\frac{\vfwhm}{\kms}},
\end{equation}
which can be obtained by integrating Equation (\ref{eq:abs}) over the LOS. Here
$\vfwhm=2\sqrt{\ln(2)}\Delta v$ is the full width at half maximum of the line
profile. The multivariate analysis performed by \citet{2003ApJ...586.1067H}
concludes that neither the historical $\taupeak$-$\Tspeak$ relationship nor the
$\taupeak$-$\Nsim$-$\Tspeak$ relationship has physical significance beyond the
relation in Equation (\ref{eq:Ttau}).

\begin{figure} 
\epsscale{0.8}
\plotone{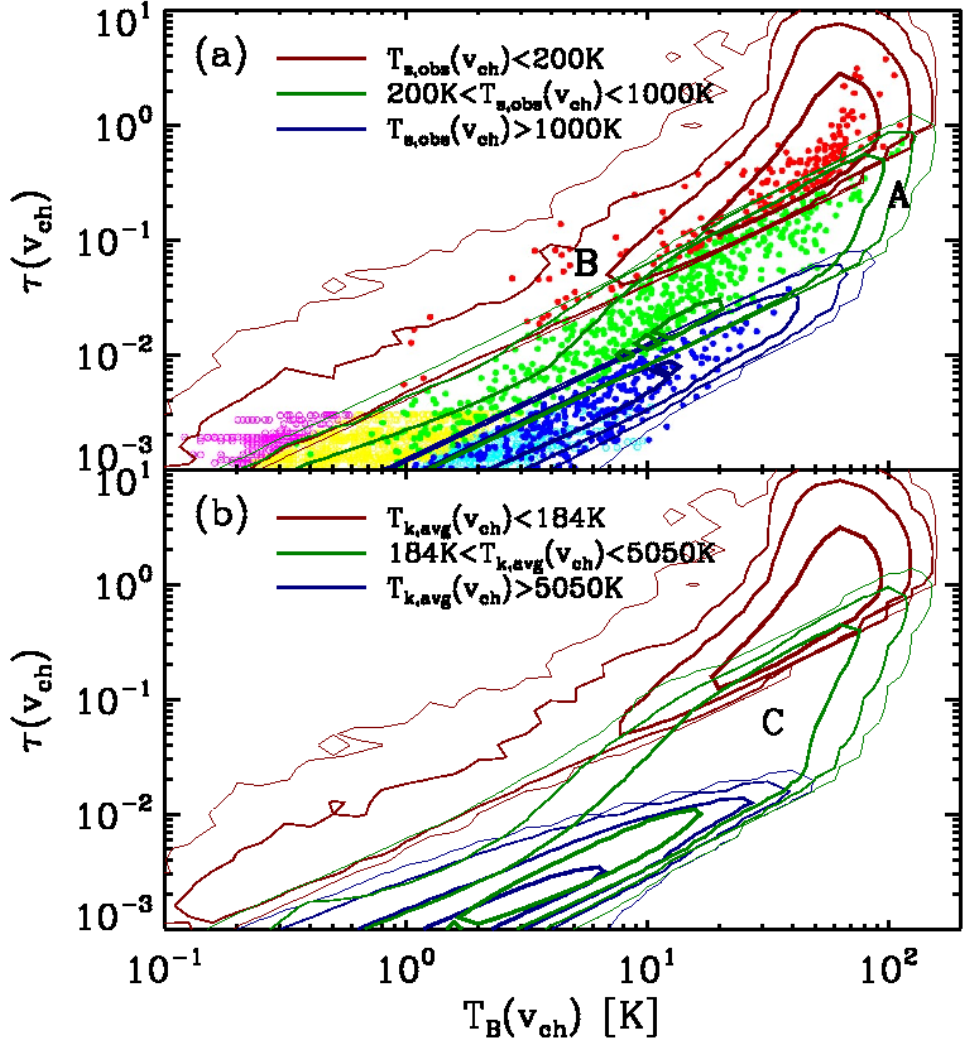}
\caption{
Distribution of the synthetic line data from simulation in the plane of the
brightness temperature $T_B(\vch)$ and the optical depth $\tau(\vch)$ for all
LOSs, channels, and snapshots of the QA10 model. The contour levels from
outside (thinner) to inside (thicker) correspond to fractions of $10^{-5}$,
$10^{-4}$, $10^{-3}$, $10^{-2}$ in each temperature range.  The red, green, and
blue contours respectively denote different ranges of (a) spin temperature
$\Tsobs$ and (b) kinetic temperature $\Tkavg$, as shown in the upper-left
corner of each panel. The underlying data in (a) is from Figure~4 of
\citet{2013MNRAS.436.2352R}.
The red, green, and blue symbols denote observed spin temperature $T_s<200\Kel$,
$200\Kel<T_s<1000\Kel$, and $1000\Kel<T_s$, respectively, for emission-detected
channels. The magenta, yellow, and cyan symbols respectively denote the same
temperature ranges for channels with non-detection.
 In (a), the observed data for emission-detected
channels (filled circles) are very well enveloped by contours drawn from the
synthetic line data: low $\Tsobs$ gas lies at high $\tau(\vch)$ and
$T_B(\vch)$, while gas with higher $\Tsobs$ lies at lower $\tau(\vch)$ and
$T_B(\vch)$. From (b), the region of high $T_B(\vch)$ and $\tau(\vch)$
corresponds to cold gas, and intermediate $T_B(\vch)$ and $\tau(\vch)$ to
intermediate-temperature gas (as in region ``C''), but gas at the lowest
$T_B(\vch)$ and $\tau(\vch)$ can be either intermediate-temperature or warm.
\label{fig:TBtau}}
\end{figure}

\begin{figure} 
\epsscale{0.8}
\plotone{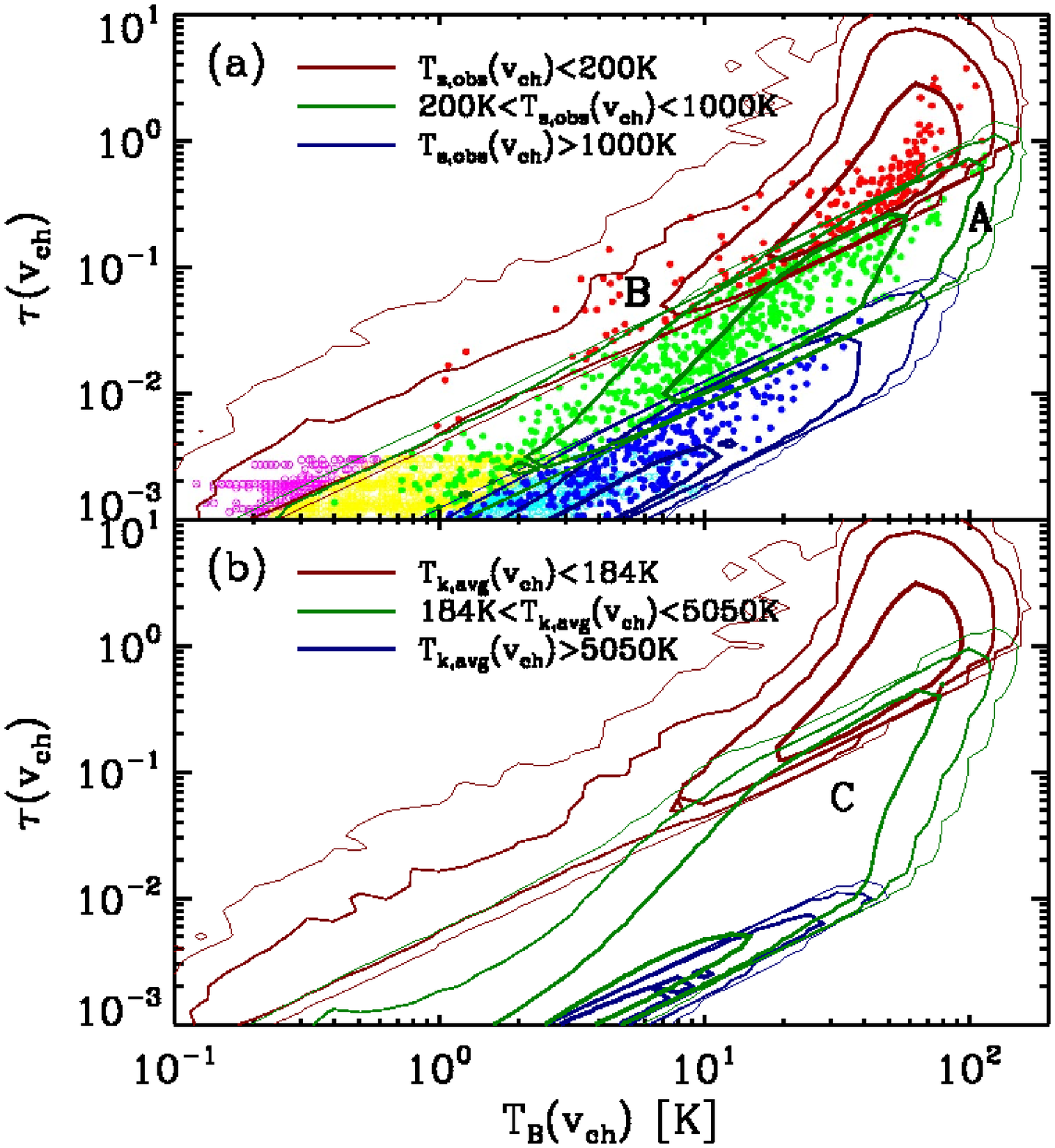}
\caption{
Same as Figure~\ref{fig:TBtau} but with the WF effect.  In (a), overall
distribution of the synthetic line data from simulation with respect to given
spin temperature ranges are similar to Figure~\ref{fig:TBtau_WF}(a). In (b),
the WNM (blue contours) is distributed in narrower region with lower
$\tau(\vch)$ since the WF effect results in higher spin temperature
(Figure~\ref{fig:Tspin}(c) and (d)).
\label{fig:TBtau_WF}}
\end{figure}

It is more informative to consider the distribution in three dimensional space
of $T_B(\vch)$-$\tau(\vch)$-$\Tsobs$ as in Figure 4 of
\citet{2013MNRAS.436.2352R}.  Figures~\ref{fig:TBtau} and \ref{fig:TBtau_WF}
display distributions of the synthetic line data without and with the WF
effect, respectively, in the plane of $T_B(\vch)$ and $\tau(\vch)$ for all
LOSs, channels, and simulation snapshots of model QA10 with $d\vch=1\kms$.
Different color contours in Figures~\ref{fig:TBtau}(a) and
\ref{fig:TBtau_WF}(a) represent different spin temperature ranges:
$\Tsobs<200\Kel$ in red, $200\Kel<\Tsobs<1000\Kel$ in green, and
$1000\Kel<\Tsobs$ in blue.  Overlaid are the observational data points
extracted from Figure~4 of \citet{2013MNRAS.436.2352R}. The distribution from
our simulation follows the observed data points very well.
\citet{2013MNRAS.436.2352R} pointed out the lack of data points for low or
intermediate optical depth measurements ($0.01<\tau<0.1$) in high
($T_B>50\Kel$; regime A) and intermediate ($1\Kel<T_B<10\Kel$; regime B)
brightness temperatures; these regions are also weakly populated by our
synthetic line data. 

To understand the paucity of data in the A regime, let us consider a LOS that
consists only of static WNM. From Equation (\ref{eq:Ttau}), we obtain
$\Tspeak\taupeak=52\Kel (\Nsim/10^{20}\psc)(\vfwhm/\kms)^{-1}$, and
$\Nsim>10^{21}\psc$ is required for $\vfwhm=10\kms$ to have $T_B\sim
\Tspeak\taupeak>50\Kel$ . Since the WNM number density is about $0.3\pcc$ in
the solar neighborhood \citep{1995ApJ...443..152W}, the path length would need
to exceed $1\kpc$ to have $\Nsim>10^{21}\psc$. Note that this is lower limit
since we assume a ``static'' medium. To have such a large column density for
the WNM within a narrow velocity range, the path length would need to be much
longer.  It is thus highly unlikely to have no CNM along such a long path,
which would be traversing a low-$|b|$ part of the ISM.  Thus, for solar
neighborhood conditions, the brightness temperature in a given channel would be
as high as in regime A only if there are CNM clouds along the LOS. With a
contribution from the CNM, the optical depth of the channel would be higher
than that of the WNM-only LOS, and the spin temperature would be lower, moving
data points upward from regime A in Figure~\ref{fig:TBtau}. Since $T_B\sim T_s$
for an optically thick channel, which is only possible when the CNM dominates,
the maximum brightness temperature would be limited to the maximum temperature
of the CNM, $T_B\simlt200\Kel$. The majority of both real and mock data points
for high optical depth have brightness temperature close to the typical spin
temperature of the CNM, $T_B\sim T_c\sim80\Kel$.

The reason for the lack of the observational points in regime B is related to
the minimum optical depth of the CNM.  The column density of a single CNM cloud
is $\sim n_c l_c$, where $n_c$ and $l_c$ are the number density and size of the
CNM cloud.  The peak optical depth of the cloud is then
\begin{equation}\label{eq:tauc}
\tau_c=0.17\rbrackets{\frac{n_c}{10\pcc}}\rbrackets{\frac{l_c}{\pc}}
\rbrackets{\frac{100\Kel}{T_s}}\rbrackets{\frac{\kms}{\vfwhm}}.
\end{equation}
The brightness temperature of the CNM is then $T_B\sim \tau_c T_s\sim 17\Kel$
for a typical small cloud. The high peak optical depth and brightness
temperature of a single CNM cloud implies that observational points in regime B
with $T_B < 10\Kel$ and $T_s<200\Kel$ would either require a single CNM layer
thinner than $1\pc$ (e.g., tiny scale atomic structure which is not resolved in
our simulations; see \citealt{2007ASPC..365....3H} for review) or far outer
wings from multiple CNM clouds. A few data points are observed in this regime
in both real and mock observations, in contrast to the relatively sharp limit
excluding regime A.

In Figures~\ref{fig:TBtau}(b) and \ref{fig:TBtau_WF}(b), we display the
distributions of the synthetic line data without and with the WF effect,
respectively, for different ranges of the harmonic mean ``kinetic''
temperature.  The red, green, and blue contours denote CNM ($\Tkavg<184\Kel$),
UNM ($184\Kel<\Tkavg<5050\Kel$), and WNM ($5050\Kel<\Tkavg$), respectively.
$\Tkavg$ is defined analogously to Equation (\ref{eq:Tsavg}) for $T_k$.  As we
have seen in Figure~\ref{fig:fcnm} and \ref{fig:fcnm_WF}, the majority of the
UNM and WNM distributions (the innermost green and blue contours) are also
completely mixed in $T_B(\vch)-\tau(\vch)$ plane.  This is in contrast to the
sharply separated spin temperature distributions in Figures~\ref{fig:TBtau}(a)
and \ref{fig:TBtau_WF}(a). The difference between contours of $\Tsobs$ and
$\Tkavg$ again warns against naive use of the spin temperature as a proxy for
the gas phase since there is not a one-to-one correspondence between $\Tsobs$
and $\Tkavg$ at all temperatures.

Using $\Tsobs$, however, it is possible to separate the $T_B(\vch)$-$\tau(\vch)$
plane into CNM-dominated, UNM-dominated, and UNM-WNM-mixed regimes. The
CNM-dominated regime is obviously defined by $\Tsobs\simlt200\Kel$. As
uncertainty in the WF parameters affects the spin temperature at low pressure
(see Figure~\ref{fig:Tspin}), the upper limit of the UNM-dominated regime is
not well defined. From Figure~\ref{fig:TBtau}(b), with no WF effect, a
conservative definition of the UNM-dominated regime is
$200\Kel\simlt\Tsobs\simlt1000\Kel$ with $T_B(\vch)\simgt10\Kel$.  From
Figure~\ref{fig:TBtau_WF}(b), with WF effect at a high level included, the
UNM-dominated regime is extended to $200\Kel\simlt\Tsobs\simlt2000\Kel$. In
other words, for $\tau(\vch)>10^{-3}$ the one-to-one correspondence between
$\Tsobs$ and $\Tkavg$ persists up to $\Tsobs\simlt1000\Kel$ (without the WF
effect) and $\Tsobs\simlt2000\Kel$ (with the WF effect). Note that green points
in the observational data are mostly in the UNM-dominated regime regardless of
the WF effect.  Therefore, the presence of the UNM is evident in the observed
data, although the relative proportions of UNM and WNM in observed data is
uncertain since the majority of the UNM is expected to be buried in the
UNM-WNM-mixed regime.

\subsection{Column Density of WNM-only LOSs}

Recently, \citet{2011ApJ...737L..33K} have proposed the existence of an
\ion{H}{1} column density threshold, $\Nth=2\times10^{20}\psc$ for CNM to form,
based on 21~cm emission and absorption line observations detailed in
\citet{2013MNRAS.436.2352R}.  They found that observed LOSs have a median spin
temperature of $\Tsim\sim 240\Kel$ for $\Nsim >\Nth$ and $\Tsim>1000\Kel$ for
$\Nsim\simlt\Nth$. This implies that the LOSs with $\Nsim>\Nth$ consist of both
CNM and WNM, while the WNM is predominant for the LOSs with $\Nsim\simlt\Nth$.

Here, we explain the observed CNM threshold behavior in the context of thermal
and dynamical equilibrium in the ISM.  As summarized in Section~\ref{sec:sim},
our model disks are in thermal and dynamical equilibrium in an average sense
\citep[][Paper~I]{2011ApJ...743...25K}.  Vertical dynamical equilibrium demands
a balance between vertical pressure support and the weight of the gas.  Also,
thermal equilibrium for a two-phase neutral medium implies the midplane thermal
pressure $\Pth$ lies between $\Pmin$ and $\Pmax$ ($\Pth\sim\sqrt{\Pmin\Pmax}$
is a good first order approximation as in
\citealt{2003ApJ...587..278W,2010ApJ...721..975O}).  The equilibrium profile
for a WNM-only vertical LOS can be approximated by a Gaussian profile with the
midplane density of $n_w(0)=\Pth/(1.1kT_w)$ and scale height of
$H_w=\sigma_{z,w}/(4\pi G\rhosd)^{1/2}$, where $T_w$ is the thermal equilibrium
temperature at $\Pth$ for the WNM, $\sigma_{z,w}=(v_{\rm turb}^2+c_w^2)^{1/2}$
is the total (thermal and turbulent) vertical velocity dispersion of the WNM,
and $\rhosd$ is the midplane density of stars and dark matter (which dominate
the potential in the solar neighborhood).  The one-sided WNM column density
along a vertical LOS is
\begin{eqnarray}
\Nwnm&=&\int_0^{\infty} n_w(z)dz=\sqrt{\frac{\pi}{2}}n_w(0)H_w
=1.14\frac{\Pth H_w}{kT_w}\nonumber\\
&=&2.0\times10^{20}\psc\rbrackets{\frac{\Pth/k}{3000\Punit}}
\rbrackets{\frac{T_w}{7000\Kel}}^{-1}
\rbrackets{\frac{\sigma_{z,w}}{10\kms}}
\rbrackets{\frac{\rhosd}{0.1\rhounit}}^{-1/2}.
\label{eq:Nwnm}
\end{eqnarray}
The reference values in the last expression are typical of the solar
neighborhood; this gives $\Nwnm$ comparable to the $\Nth$ value reported in
\citet{2011ApJ...737L..33K}.  In Column (7) of Table~\ref{tbl}, we list
corresponding $\Nwnm$ calculated from the numerical outcomes of $\Pth$ and
$H_w$ in several numerical models representing a range of galactic conditions.
For arbitrary galactic latitude $b$, the column density of a WNM-only LOS would
be $\Nwnm/\sin|b|$.

\begin{figure} 
\epsscale{1.0}
\plotone{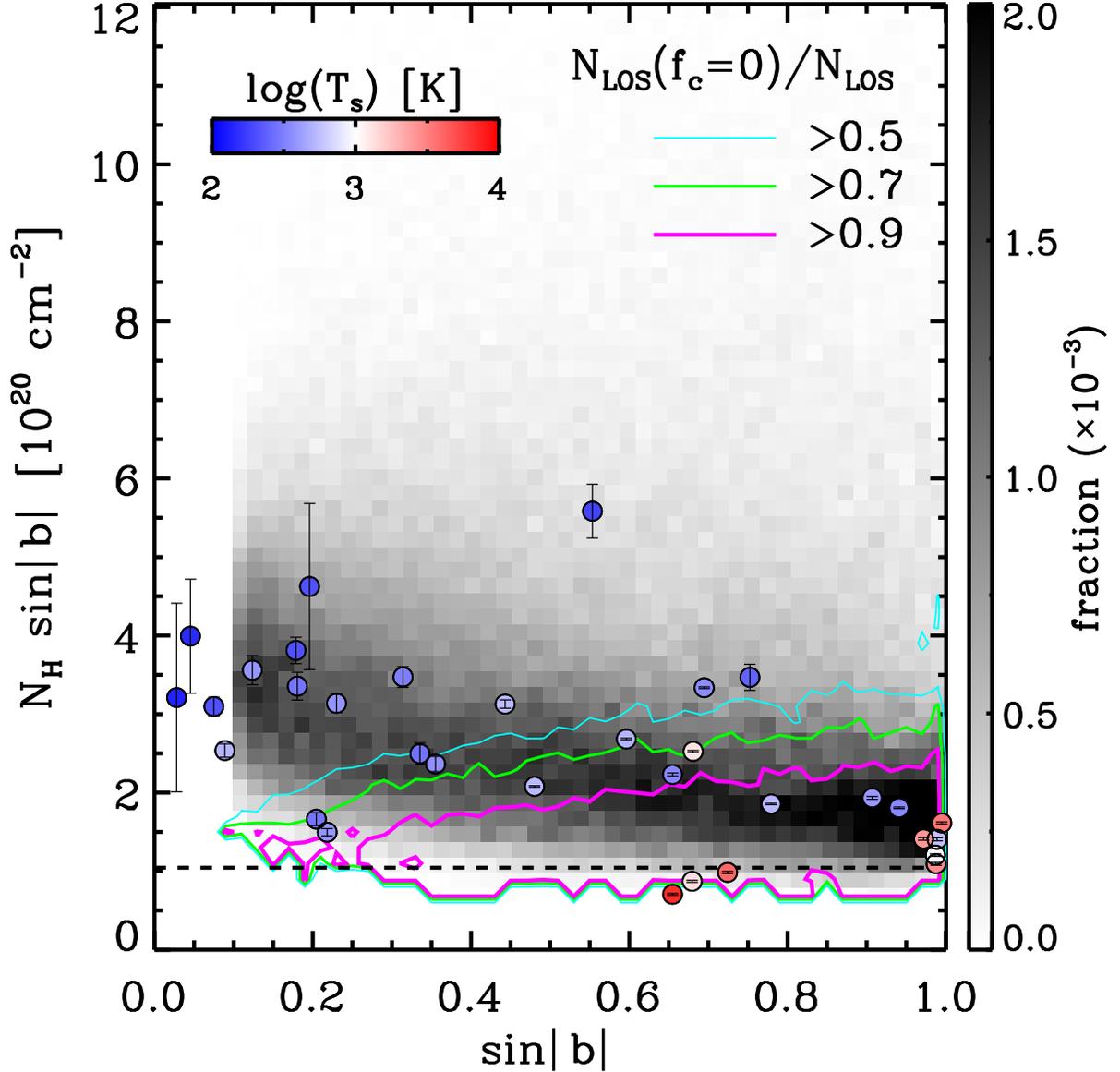}
\caption{
Distribution of the mock observation data of the column density (from model
QA10) projected to the vertical LOS $\Nsim \sin|b|$ as a function of $\sin|b|$
for all LOSs and simulation snapshots.  The grey scale displays the
number fraction in each bin, with $\Nsim \sin|b|$ increasing at small $|b|$
because cold gas is concentrated more towards the midplane than warm gas.  The
contours indicate the fractions of the mock observation data in each bin that
is no-CNM ($f_c=0$) increasing from 0.5 (cyan) to 0.7 (green) to 0.9 (magenta).
The symbols are observed data points from \citet{2013MNRAS.436.2352R}; the
color of each symbol represents the logarithm of the spin temperature.  The
column density of the WNM-only LOS (see Equation (\ref{eq:Nwnm})), $\Nwnm$
(using $\Pth$ and $H_w$ for model QA10 from Table~\ref{tbl}), is shown in the
dashed line.}\label{fig:Nwnm}
\end{figure}

Figure~\ref{fig:Nwnm} displays the distribution of $\Nsim \sin|b|$ in grey
scale as a function of $\sin|b|$, drawn from our model QA10.  The observed data
points, plotted as filled circles with color denoting the spin temperature
(taken from \citealt{2013MNRAS.436.2352R}) follow the same general distribution
as the mock observation data (grey scale).  In order to quantify where the WNM
dominates, we calculate the fraction of no-CNM LOS ($f_c=0$) in each bin. From
outside to inside, the contours demark no-CNM LOS fraction of $>50\%$ (cyan),
$>70\%$ (green), and $>90\%$ (magenta).  The innermost magenta contour (more
than 90\% of LOSs lack CNM) envelopes the observed data points with high spin
temperatures ($>10^3\Kel$). The horizontal dashed line marks $\Nwnm$ from
Equation (\ref{eq:Nwnm}). Note that the spin temperatures of the observed data
change systematically from small (CNM dominated) to large (UNM+WNM dominated)
as $\Nsim \sin|b|$ becomes smaller. This is consistent with
Figures~\ref{fig:fcnm} and \ref{fig:fcnm_WF}, which show $\log\Tsyn>3$ only for
LOSs without CNM.

Since the disk is highly turbulent and time-variable, the midplane pressure of
the WNM can be somewhat larger and smaller than the mean midplane pressure
$\Pth$.  Thus, it is possible to have no-CNM LOSs with column density larger
than $\Nwnm/\sin|b|$. However, most LOSs, especially at small $|b|$, consist of
both CNM and WNM, as seen in the region excluded from the contours in
Figure~\ref{fig:Nwnm}. Where both WNM and CNM are present, $\Nsim\sin|b|$ will
exceed $\Nwnm$. We note, however, that the lowest values of $\Nsim\sin|b|$ from
\citet{2013MNRAS.436.2352R} appear in the regime that we expect will be
WNM-dominated. We therefore suggest that the ``cold threshold'' column density
seen by \citet{2011ApJ...737L..33K} is not the manifestation of a minimum
shielding column, but instead represents the vertical \ion{H}{1} column with
only WNM that is consistent with both thermal and dynamical equilibrium in the
local Milky Way disk.

\begin{figure} 
\epsscale{1.0}
\plotone{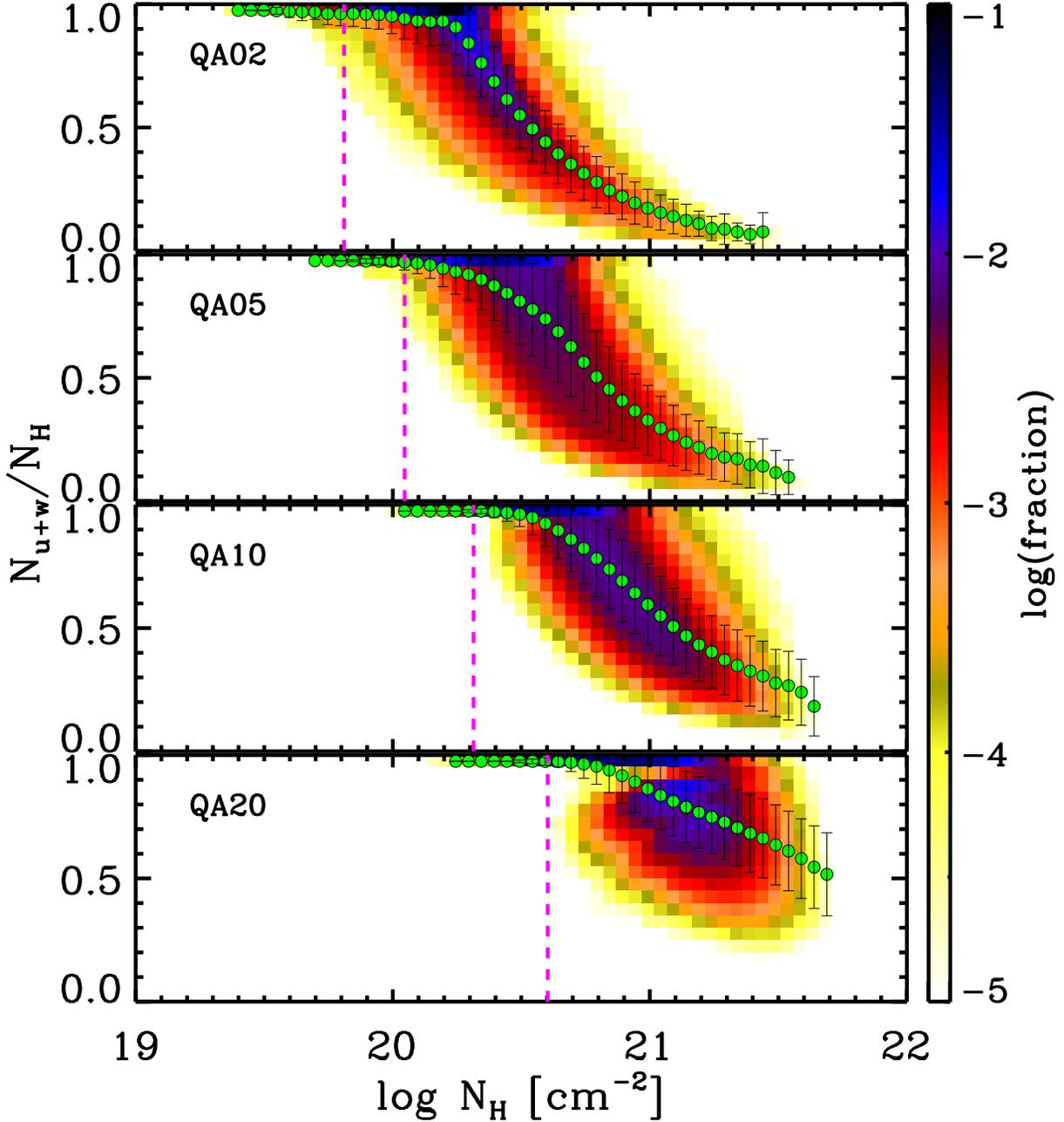}
\caption{
Distribution of the UNM+WNM mass fraction $N_{u+w}/\Nsim$ as a function of
total column density $\Nsim$ for all vertical LOSs at all times in each
simulation, for different galactic disk model environments (see
Table~\ref{tbl}). Here the column density is defined by integrating over the
entire disk vertically in contrast to the half-thickness integration of
Equation (\ref{eq:Nwnm}).  The mean and standard deviation of the distribution
in each $\Nsim$ bin are shown as symbols and errorbars, respectively.  The
magenta dashed line shows the predicted value $2\Nwnm$ for a WNM-only vertical
column in thermal and dynamical equilibrium (see text). The decline below
near-unity of the UNM+WNM mass fraction occurs at $\sim2\Nwnm$, as expected.}
\label{fig:ntau_ext}
\end{figure}

If external galaxies were observed with sufficiently high resolution and
sensitivity, the column density of a WNM-only LOS would similarly vary with
local disk conditions.  In order to address this, we consider what an
\ion{H}{1} observer would see for different extragalactic ISM conditions as
modeled by the set of simulations listed in Table~\ref{tbl}.  We vertically
integrate the gas density through the entire disk to obtain the column density
$\Nsim\equiv\int_{-\infty}^{\infty} n dz$.  Note that this column density is
twice that defined by Equation (\ref{eq:NHsim}), which assumes an observer at
the midplane.  We also calculate the column density of the CNM ($N_c$) and the
UNM \emph{plus} WNM ($N_{u+w}$) in the same way.  Here, we consider the UNM and
WNM together since they are not distinguishable in \ion{H}{1} line observations
(see Figures~\ref{fig:fcnm} and \ref{fig:TBtau}).

Figure~\ref{fig:ntau_ext} displays the distribution of the UNM+WNM mass
fraction $N_{u+w}/\Nsim$ as a function of the total column density $\Nsim$.
The color scale shows the distribution of the logarithmic number fraction for
each of the simulations listed in Table~\ref{tbl}. The symbols and errorbars
respectively plot the mean and standard deviation of the distribution for each
$\Nsim$ bin.  The vertical dashed line in each panel denotes the column density
$2\Nwnm$ for a WNM-only LOS under equilibrium conditions through the full disk
thickness.  For $\Nsim < 2\Nwnm$, most of LOSs have $N_{u+w}/\Nsim$ higher than
$90\%$.  Above this column density, LOSs are dominated by the two-phase
mixture. Thus, $2\Nwnm$ represents a lower limit on the total column for cold
gas to be present.

The overall distribution and the threshold column density move toward higher
$\Nsim$, as the total gas surface density increases from $2.5\Surf$ to
$20\Surf$ (from QA02 to QA20 models). The self-regulated equilibrium model for
atomic-dominated regions \citep{2010ApJ...721..975O} predicts
$\Pth\propto\SigSFR\propto\Sigma\sqrt{\rhosd}$ and $H_w\propto 1/\sqrt{\rhosd}$
if the ratio of thermal to total pressure, and the vertical velocity dispersion
are approximately constant, as verified numerically for QA-series
\citep[][Paper~I]{2011ApJ...743...25K}. From Equation (\ref{eq:Nwnm}), we thus
expect $\Nwnm\propto\Pth H_w\propto \Sigma$, explaining why $\Nwnm$ increases
nearly linearly in the QA Series (see Table~\ref{tbl}).

\section{Summary and Discussion}

The very first step to deduce the \ion{H}{1} column density from 21~cm line
observations is to convert the observed brightness temperature $T_B(l,b,\vch)$
to channel column density using Equation (\ref{eq:Nsyn}) or (\ref{eq:Nthin}).
Despite the importance of this first conversion step, the validity and
uncertainty of the conversion methods have not previously been tested and
quantified with realistic ISM models.  At a minimum, a realistic ISM model
should include multi-scale turbulence, as well as self-consistent density and
temperature structure responding to the heating and cooling of the dynamic ISM.
Our recent ISM simulations in Paper~I include these ingredients, and thus
provide a valuable testbed for evaluating 21~cm diagnostic techniques.  Since
our simulations are local, we limit our analysis to latitudes larger than
$|b|>5^\circ$ in which horizontal variations of the ISM play a lesser role.

Our main findings are summarized as follows.

\begin{enumerate}

\item By conducting mock observations toward random LOSs, we find that the
observed spin temperature $\Tsobs$ deduced from the brightness temperature
and the optical depth (see Equation (\ref{eq:Tsobs})) agrees very well with the
true harmonic mean spin temperature $\Tsavg$ (see Equation (\ref{eq:Tsavg})).
The agreement is within a factor of 1.5 even for the channel optical depths as
large as $\tau(\vch)\sim10$.  Since $\Tsobs$ effectively assumes a single
component along the LOS, this agreement implies that for the adopted velocity
channel width of $1\kms$, there is limited LOS overlap of opaque CNM clouds. 
The agreement $\Tsavg\approx\Tsobs$ in individual velocity channels also leads
to good agreement between ``observed'' and ``true'' velocity-integrated
properties, including $\Nsyn\approx\Nsim$ and $\Tsyn\approx\Tsim$ (see
Figure~\ref{fig:NHcomp}).  $\Nsyn$ (see Equation \ref{eq:Nsyn}) is also known
as the ``isothermal'' estimator of the \ion{H}{1} column density  and widely
used in observations \citep{1982AJ.....87..278D,2013MNRAS.432.3074C}.  The
``thin'' column density is within a factor of $\sim0.7$ of the ``true'' column
density for $\tauint<10$, comparable to the observed ratio of ``thin'' to
opacity-corrected column density $\sim 0.6-0.8$ \citep{2003ApJ...585..801D}.

\item 
In our analysis, we calculate the spin temperature with and without the WF
effect via the Ly-$\alpha$ resonant scattering, which provide upper and lower
limits for the spin temperature of the WNM, respectively.  As a consequence, we
find the harmonic mean spin temperature (Equation \ref{eq:Tsyn}) is limited to
$\Tsyn\simlt2000\Kel$ without the WF effect (see Figure~\ref{fig:fcnm}) and
$\Tsyn\simlt4000\Kel$ with the WF effect (see Figure~\ref{fig:fcnm_WF}). There
are a few absorption line observations of the WNM that have reported spin
temperature up to $\sim 6000\Kel$
\citep{1998ApJ...502L..79C,2003MNRAS.346L..57K}.
The possible underestimation of the WNM spin temperature in our analysis (we
omit collisional transitions due to electrons) might lead to higher optical
depth of the WNM than in the real ISM. However, the optical depth is already
quite small in the WNM, and we find a similar distribution in the space of
$T_B(\vch)$-$\tau(\vch)$-$\Tsobs$ to observations irrespective of the method
for the spin temperature calculation (see Figures~\ref{fig:TBtau} and
\ref{fig:TBtau_WF}).

\item Our analysis shows that thermally-unstable and true warm gases appear in
comparable proportions along all LOSs for most values of $\Tsyn$
(Figures~\ref{fig:fcnm} and \ref{fig:fcnm_WF}). However, we do find that for
$\Tsyn$ in the range $\sim 500\Kel-1000\Kel$, UNM dominates over WNM.  The
detection of absorption with spin temperature in a range of $\Tsyn\sim
500-5000\Kel$
\citep[e.g.,][]{1998ApJ...502L..79C,2002ApJ...567..940D,2003MNRAS.346L..57K,2010ApJ...725.1779B}
may imply the possible existence of thermally unstable gas.  More definitive
evidence for the UNM is given by a distribution of observational data in the
space of $T_B(\vch)$-$\tau(\vch)$-$\Tsobs$ (Figures~\ref{fig:TBtau} and
\ref{fig:TBtau_WF}). 
We show that the UNM alone populates the regime with
$200\Kel\simlt\Tsobs\simlt1000\Kel$ (near point ``C'' in
Figure~\ref{fig:TBtau}(b)) without the WF effect. With the WF effect, this
regime can be extended to $200\Kel\simlt\Tsobs\simlt2000\Kel$ (see
Figure~\ref{fig:TBtau_WF}(b)). This is because the one-to-one correspondence
between $\Tsobs$ and $\Tkavg$ persists up to at least $\Tsobs\simlt1000\Kel$
and at best $\Tsobs\simlt2000\Kel$ without and with WF effect, respectively.
The presence of observational data points in this region strongly suggest that
the presence of thermally unstable gas.  Since the majority of the UNM and WNM
occupy the low optical depth regime $\tau(\vch)\simlt10^{-2}$, with high spin
temperature $\Tsobs\simgt1000\Kel$ where their distributions are completely mixed,
however, the exact mass fractions of the UNM and WNM are difficult to derive
from \ion{H}{1} 21~cm observations. 

\item While the spin temperature provides only limited ability to differentiate
UNM and WNM, it is a very good probe of the CNM mass fraction.  From Equation
(\ref{eq:fcnm}), the CNM mass fraction for moderate $\Tsyn$ is $\fcsyn\approx
T_c/\Tsyn$. In our simulations, the median CNM temperature is $80\Kel$, and
Figure~\ref{fig:fcnm}(a) shows that for LOSs with low spin temperature
$\Tsyn\simlt400\Kel$, $\fcsyn\approx 80\Kel/\Tsyn$ fits quite well.  Because CNM
temperature is not a single constant, however, there is an inherent uncertainty
of about a factor of 2 in this result.  Using Gaussian decomposition of
emission/absorption lines, \citet{2003ApJ...585..801D} have found the CNM
temperature are in range of $40\Kel\simlt T_c\simlt100\Kel$ with median value of
$\sim65\Kel$.  \citet{2003ApJ...586.1067H} also have reported similar
distributions with median and mass-weighted CNM temperatures of $48\Kel$ and
$70\Kel$, respectively. 

Interesting results for the CNM mass fraction have recently been derived by
emission/absorption line pairs from galactic plane surveys
\citep{2003AJ....125.3145T,2005ApJS..158..178M,2006AJ....132.1158S}.
\citet{2009ApJ...693.1250D} have shown that the radial dependence of harmonic
mean spin temperature in the Milky Way is nearly flat, implying that nearly
constant CNM mass fraction out to $25\kpc$. Our numerical models (Paper~I; see
also \citealt{2011ApJ...743...25K}) indeed find that mean CNM mass fraction
varies only from $\sim0.2$ to $0.4$ for a wide range of conditions.

\item For a given equilibrium disk condition with midplane thermal pressure
$\Pth$ and scale height of the WNM $H_w$, there is a maximum WNM-only vertical
column density $\Nwnm$ (Equation (\ref{eq:Nwnm})).  This is comparable to the
observed column density where a transition in $\Tsyn$ occurs,
$\Nth\sim2\times10^{20}\psc$ \citep{2011ApJ...737L..33K}.  The detailed
distributions of our mock observations shows that LOSs with projected column
density $\Nsim\sin|b|$ smaller than $\Nwnm$ are highly likely to consist only
of the WNM. $\Nwnm$ therefore represents well the transition column density
from warm-dominated LOSs to LOSs with a two-phase mixture, for a wide range of
model parameters (Figure~\ref{fig:ntau_ext}).
\end{enumerate}

\acknowledgements{We are grateful to the referee for an extremely helpful
report, including encouragement to expand our discussion of the Wouthuyen-Field
effect. This work was supported by grant AST0908185 from the National Science
Foundation.  The work of W.-T. K. was supported by the National Research
Foundation of Korea (NRF) grant funded by the Korean government (MEST), No.
2010-0000712.  The simulations used in this paper were performed by the
facilities of the Shared Hierarchical Academic Research Computing Network
(SHARCNET:www.sharcnet.ca) and Compute/Calcul Canada. }

\end{document}